\advance\hoffset by 5mm
\newif\ifPDFLaTeX
\expandafter\ifx\csname pdfoutput\endcsname\relax\else\PDFLaTeXtrue\fi
\documentclass[epj]{svjour}%
\let\makeheadbox\relax
\usepackage{amsmath,amssymb}
\usepackage{url}
\usepackage{graphicx}
\usepackage{thophys}

\newcommand{\ii}{\mathrm{i}}

\def\MAKE/{\texttt{make}}
\def\MAKEFILE/{\texttt{Makefile}}
\def\PERL/{\texttt{PERL}}
\def\OCAML/{\texttt{O'Caml}}
\def\FORTRAN/{\texttt{Fortran}}
\def\FORTRANNINETY/{\texttt{Fortran\,90}}
\def\FORTRANNINETYFIVE/{\texttt{Fortran\,95}}
\def\FORTRANOBJECTS/{\texttt{Fortran\,2003}}
\def\FORTRANSEVENTYSEVEN/{\texttt{FORTRAN77}}
\def\C/{\texttt{C}}
\def\Cpp/{\texttt{C++}}
\def\JAVA/{\texttt{JAVA}}
\def\WHIZARD/{\texttt{WHIZARD}}
\def\OMEGA/{\texttt{O'Mega}}
\def\OMEGALIB/{\texttt{omegalib}}
\def\COMPHEP/{\texttt{CompHEP}}
\def\GRACE/{\texttt{GRACE}}
\def\MADGRAPH/{\texttt{MadGraph}}
\def\MADEVENT/{\texttt{MadEvent}}
\def\HELAS/{\texttt{HELAS}}
\def\HELAC/{\texttt{HELAC}}
\def\ALPHA/{\texttt{ALPHA}}
\def\SHERPA/{\texttt{Sherpa}}
\def\CIRCE/{\texttt{CIRCE}}
\def\LUSIFER/{\texttt{LUSIFER}}
\def\GUINEAPIG/{\texttt{GuineaPig}}
\def\PDFLIB/{\texttt{PDFlib}}
\def\LHAPDF/{\texttt{LHAPDF}}
\def\VAMP/{\texttt{VAMP}}
\def\VEGAS/{\texttt{VEGAS}}
\def\FEYNMF/{\texttt{FeynMF}}
\def\GAMELAN/{\texttt{gamelan}}
\def\METAPOST/{\texttt{METAPOST}}
\def\POSTSCRIPT/{\texttt{PostScript}}
\def\STDHEP/{\texttt{STDHEP}}
\def\PYTHIA/{\texttt{PYTHIA}}
\def\HERWIG/{\texttt{HERWIG}}
\def\LANHEP/{\texttt{LanHEP}}
\def\HEPMC/{\texttt{HepMC}}
\def\LHEF/{\texttt{LHEF}}
\def\SINDARIN/{\texttt{SINDARIN}}
\def\FEYNRULES/{\texttt{FeynRules}}
\onecolumn
\begin{document}
\title{%
  \texttt{WHIZARD} -- 
  Simulating Multi-Particle Processes
  at LHC and ILC}
\author{%
  Wolfgang
  Kilian\thanks{\email{kilian@hep.physik.uni-siegen.de}}\inst{,1}\and
  Thorsten
  Ohl\thanks{\email{ohl@physik.uni-wuerzburg.de}}\inst{,2}\and
  J\"urgen Reuter\thanks{\email{juergen.reuter@desy.de}}\inst{,3,4,5}}

\institute{%
    Department Physik, University of Siegen,
    D--57068 Siegen, Germany \and
    Institut f\"ur Theoretische Physik und Astrophysik,
    University of W\"urzburg,\\ D--97074 W\"urzburg, Germany \and
    Theory Group, DESY Hamburg, D--22603 Hamburg, Germany \and
    School of Physics, University of Edinburgh, 
    Edinburgh EH9 3JZ, Scotland \and
    Physikalisches Institut, University of Freiburg,
    D--79104 Freiburg, Germany}
\date{January 2011}
\headnote{\begin{footnotesize}DESY 11-126, \quad EDINBURGH-2010-36, \quad
    FR-PHENO-2010-037, \quad SI-HEP-2010-18
     \end{footnotesize}}
\abstract{%
   We describe the universal Monte-Carlo (parton-level) event
   generator
   \texttt{WHIZARD}\footnote{\url{http://whizard.event-generator.org}},
     version 2. 
  The program automatically computes complete tree-level matrix
  elements, integrates them over phase space, evaluates distributions
  of observables, and generates unweighted partonic event samples.
  These are showered and hadronized by calling external codes,
  either automatically from within the program or via standard
  interfaces. There is no conceptual limit on the process
  complexity; 
  using current hardware, the program has successfully been applied to
  hard scattering processes with up to eight particles in the
  final state.  Matrix elements are computed as helicity
  amplitudes, so spin and color correlations are retained.  For event
  generation, processes can be concatenated with full spin correlation, so
  factorized approximations to cascade decays are possible when complete
  matrix elements are not desired.
  The
  Standard Model, the MSSM, and many alternative models such as
  Little Higgs, anomalous couplings, or effects of extra dimensions or
  noncommutative SM extensions have been implemented.
  Using standard interfaces to parton shower and
  hadronization programs, \texttt{WHIZARD}
  covers physics at hadron, lepton, and photon colliders.
  \PACS{%
    {11.15.Bt}{General properties of perturbation theory} \and
    {11.15.-q}{Gauge Field Theories} \and
    {11.80.Cr}{Kinematical properties} \and
    {12.38.Bx}{Perturbative calculations}}}
\maketitle
\tableofcontents
\section{The Need for Multi-Particle Event Generators}

At the LHC and the future ILC experiments, we hope to uncover the
mechanism of electroweak symmetry breaking and to find signals of physics
beyond the Standard Model (SM).  Many of the key elementary processes
that have to be investigated for this purpose -- such as weak-boson
fusion and scattering, $t\bar tH$ production, supersymmetric cascades,
exotica -- are much more complex than the SM processes that were
accessible at previous colliders.  Simultaneously, the requirements
for theoretical predictions of ILC processes will significantly
surpass the LEP precision, while LHC data can only be meaningfully
analyzed if a plethora of SM and possibly non-SM background and
radiation effects are theoretically under control.

Monte-Carlo tools such as \PYTHIA/~\cite{pythia} or
\HERWIG/~\cite{herwig} are able to predict signal rates for SM as well
as various new-physics processes.  These programs contain hard-coded
libraries of leading-order on-shell matrix elements for simple
elementary scattering, decay, and radiation processes.  However, the
requirements of precision and background reduction will only be
satisfied if Monte-Carlo simulation programs can correctly handle
off-shell amplitudes, multi-particle elementary processes, dominant
radiative corrections, and matrix-element/parton-shower matching.
While previously the main difficulty in Monte-Carlo
simulation was the proper description of showering and
non-perturbative QCD, more recent codes also address the technical
problems of partonic multi-particle simulation without on-shell
factorization approximations.
The variety and complexity of proposed new-physics models
makes it impractical to code every single process in a common library.
There is obvious need for automated and flexible tools for
multi-particle event generation, capable of simulating all kinds of
physics in and beyond the SM. 

This field has been pioneered by the \COMPHEP/~\cite{comphep} and
\GRACE/~\cite{grace} collaborations, for processes of still limited
complexity.  The \MADGRAPH/~\cite{madgraph} amplitude generator for
the \HELAS/~\cite{HELAS} library provided the first automatic tool for
computing multi-particle amplitudes.  In the last decade, the rapid
increase in computing power together with the development of
multi-channel integration
techniques~\cite{Kleiss:1994qy,vamp,madevent} has made multi-particle
phase space accessible to Monte-Carlo simulation.  Furthermore, new
ideas~\cite{mc@nlo,Catani:2001cc} have opened the path for a
consistent inclusion of higher-order QCD effects in the simulation.

Consequently, several new approaches to the problem of realistic and
universal physics simulation at the LHC and ILC have been
implemented~\cite{alpha,whizard,helac,madevent,sherpa}. In 
this paper, we describe the current status of the
\WHIZARD/~\cite{whizard} package, which provides a
particular approach to the challenges of multi-parton matrix-element
construction and event generation in multi-particle partonic
phase-space.  Its main components are
the \OMEGA/~\cite{omega,omega2} matrix element generator, the
\VAMP/~\cite{vamp} adaptive multi-channel multi-dimensional
integration library, 
and its own module for constructing
suitable phase-space parameterizations.  These parts, which use
original algorithms and implementations, are the focus of
the present paper.

For physics event simulation, \WHIZARD/ offers several
possibilities.  It implements the standard Les Houches interface, so
shower and hadronization codes can be externally attached.
Alternatively, \WHIZARD/ can perform showering and hadronization by
internally calling PYTHIA with proper matching, so in this mode it behaves 
as a complete tree-level event generator for collider
physics.  A third path, which is not yet in
production status and will be the subject of a separate
publication~\cite{whizard-shower},
involves an independent parton-shower module that is to be
combined with (external) hadronization.

\section{Physics Simulation with \WHIZARD/}

\subsection{Purpose and Scope}

\WHIZARD/ is a program to compute cross sections and distributions of
observables, and to generate simulated event samples, for
hard scattering and decay processes of particles at
high-energy colliders.  The theoretical framework is set by
leading-order perturbation theory, i.e., tree-level partonic
matrix elements.
These are calculated in a fully automatic way.  Refinements such as
higher orders in perturbation theory, form factors, or other
non-perturbative effects in the hard-scattering process can
be added manually.

The physics described by \WHIZARD/ is given by the Standard Model of
strong and electroweak interactions, and by well-known extensions of
it, such as the minimal supersymmetric Standard Model (MSSM), Little
Higgs models, anomalous couplings, and more.

The program covers physics at all experiments in elementary particle
physics, including, for instance, LEP, Tevatron, the LHC, and the
ILC.  LHC physics is described by the convolution of parton
distribution functions with hard-scattering processes.  QCD effects
that cannot be described by fixed-order perturbation theory are
accounted for via standard interfaces to external programs. \WHIZARD/
is particularly adapted to ILC physics due to the detailed description
of beam properties (energy spread, crossing angle, beamstrahlung, ISR,
polarization)~\footnote{The ILC-specific features of \WHIZARD/~1 have
  not yet been completely re-enabled in \WHIZARD/~2.}.   

In contrast to programs such as \PYTHIA/~\cite{pythia} or
\HERWIG/~\cite{herwig}, \WHIZARD/ does not contain a fixed library of physics
processes, and it is not limited to a small number of particles at the hard
scattering level.  Instead, for any process that is possible at tree
level in the selected physics model, the matrix element is computed as
needed and translated into computer code, using the \OMEGA/ program.
The \OMEGA/ algorithm is designed to compute helicity amplitudes in
the most efficient way, by eliminating the redundancies in
the calculation that would show up in a naive Feynman-graph expansion.

The phase space setup is implemented in a form suitable for
efficient multi-parton integration and event generations, and the
further requirements for a complete event generator in a collider
environment can also be handled in a fully automatic way, partly
by calling external codes.  From the user's perspective, \WHIZARD/
thus has a similar purpose and scope as \COMPHEP/~\cite{comphep},
\MADEVENT/~\cite{madgraph,madevent}, and \SHERPA/~\cite{sherpa}, which
also aim at the simulation of multi-particle processes, the latter
including its own modules for non-perturbative QCD
effects. 
All mentioned codes use independent and different algorithms and
implementations, and have different ranges of applicability and
degrees of optimization.


\subsection{Workflow}

After the installation of the program as described in
sec.~\ref{sec:installation}, \WHIZARD/ is steered by a script that the user
provides, written in \SINDARIN/, a domain-specific script language
specifically designed for this task.  The script can be provided on the
command line, as a file, or distributed among several files.  Alternatively,
it can be typed and executed in interactive mode.

The language lets the user specify the physics model, scattering and decay
processes, physics parameters, and run parameters, in a simple assignment
syntax.  Tasks such as integration and simulation are executed as commands in
\SINDARIN/.  The language furthermore implements histograms and plots (via
an interface to \LaTeX\ and \METAPOST/), and it supports user-defined
variables, conditional execution of commands as well as parameter scans and
more complex workflow patterns.

In a straightforward run, the user script will select a physics model, specify
various processes within that model, declare the structure of colliding beams,
set physics parameters, define cuts, and integrate the processes.  Once the
integral (the cross section or partial width for scatterings and decays,
respectively) is known, the program is able to generate simulated events for a
process, which can be analyzed and histogrammed, or simply written to file for
processing by external programs.

The first choice selects one of the physics models that are supported by
\WHIZARD/, an overview of which can be found in sec.~\ref{sec:models}.  For
supersymmetric models in particular, an interface to the SUSY Les Houches
Accord (SLHA I+II)~\cite{slha} simplifies parameter input.  After a specific
physics model has been selected, the user can specify a list of partonic
processes for which the matrix elements are generated.  \WHIZARD/
automatically calls the matrix-element generator \OMEGA/ with appropriate
parameters for generating code, compiles the generated files, and links them
dynamically, so they are immediately available to integration and simulation.
Optionally, physics processes can be restricted to specific intermediate
states, or a class of processes can be combined by summing over equivalent
initial-state or final state particles in the process definition.

In general, \WHIZARD/ itself detects the complexity of the processes
under consideration and estimates the number of necessary random
number calls that are needed for a stable integration.  \WHIZARD/~2 does not
enforce any cuts for regulating the infrared and collinear singularities.
Instead, the \SINDARIN/ language allows for specifying rather generic cut,
trigger, and veto conditions that can be calculated from an extensible set of
partonic observables.  These conditions can be formulated both in a
process-specific or process-independent way, along the lines of an actual
(partonic) process analysis.

For analysis purposes, event generation can be switched on. There are
two options: weighted and unweighted events. The effort for
unweighting the Monte Carlo events grows with the number of external
particles, but is well under control. Weighted distributions need much
less generation time, and result in smoother distributions, but their
fluctuations do not correspond to real data with the same number of
events.  In addition, using weighted distributions in detector
simulations can exercise the detector in regions of phase space that
are thinly populated by real data, while scarcely probing regions of
phase space where most of the real events will lie. The data output is
available in several different event formats, ranging from a very long
and comprehensive debugging format to machine-optimized binary format. 
Generated events are mandatory for analysis, i.\,e.~for producing
histograms. Histograms can easily be generated with \WHIZARD/'s own
graphics package, using the same type of expressions as used for specifying
cuts, energy scale, or standardized data files can be written out.


\subsection{Program Structure}

\begin{figure}
  \begin{center}
  \includegraphics{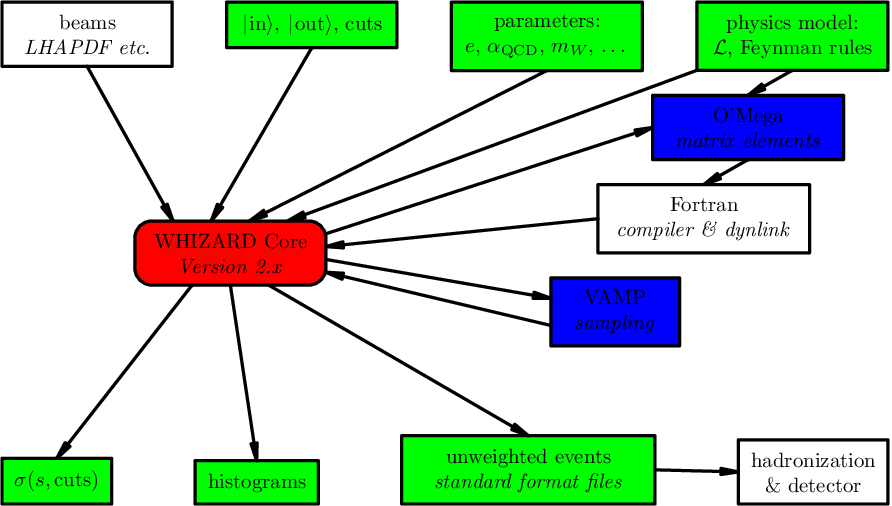}
  \end{center}
  \caption{\label{fig:structure}%
    The overall structure of \WHIZARD/.}
\end{figure}

The overall architecture of \WHIZARD/ is sketched in
figure~\ref{fig:structure}.  The structure of largely independent
software components is both good programming practice and reflects the
development history.  \WHIZARD/~\cite{whizard}, \OMEGA/~\cite{omega,omega2}
and \VAMP/~\cite{vamp} were developed independently and communicate
only via well defined interfaces.  While \OMEGA/ and \VAMP/ were
designed to solve only one problem, optimized matrix element
generation (see section~\ref{sec:omega} for details) and adaptive multi-channel
importance sampling~\cite{vamp} respectively, the \WHIZARD/ component
plays a dual r\^ole, both as phase space generator and as the central
broker for the communication among all the other components.

This component structure makes it possible to implement each component
in a programming language most suited to its purpose.\footnote{The
  choices made reflect the personal opinions of the authors on a
  subject that is often the realm of highly emotional arguments.}  (In
this context, \FORTRAN/ refers to the current standard
\FORTRANOBJECTS/.  Currently, this standard is not yet universally
adopted by compiler vendors; for this reason, the current \WHIZARD/
implementation uses a specific subset of the \FORTRANOBJECTS/ standard
that is supported by various widely available \FORTRAN/ compilers.)
\begin{itemize}

  \item \WHIZARD/ organizes data both from a physics perspective
    (implementing, e.g., quantum correlations, and phase space kinematics) and
    for the user interface (implementing, e.g., lexer, parser, and compiler
    for the \SINDARIN/ language), and it manages the interfaces to external
    programs and to the operating system.  The simultaneous requirements of
    handling complex data structures and efficiently evaluating numerical
    expressions are well met by modern programming languages such as
    \FORTRAN/ and \Cpp/.  For the \WHIZARD/~2 implementation,
    \FORTRAN/ was chosen, so the program takes advantage of
    efficient numerics, high-level memory management,
    native array support, and modular programming with data
    encapsulation.  String handling
    is done by the standard 
    \verb|iso_varying_string| module.  The operating-interface is cared for by
    dynamic procedure pointers and portable \C/ interoperability.  Furthermore,
    the \FORTRAN/ implementation allows for directly interfacing the \VAMP/
    integration library.

    \WHIZARD/~2 is written in an object-oriented programming
    style, to ensure scalability and extensibility.\footnote{As this is
    written, free \FORTRAN/ compilers do not yet implement the
    \FORTRANOBJECTS/ standard
    completely, so \WHIZARD/~2.0 had to refrain from using certain
    new syntax features.  A future revision will
    exploit these 
    features, aiming at a considerable simplification of the program text
    without altering the structure.}  Nontrivial data
    objects are allocated and deallocated dynamically, and global state
    variables are confined to few and well-defined locations.

  \item \OMEGA/ as the generator for matrix-element code has no numerical
    objectives, but is very similar to a 
    modern retargetable optimizing compiler instead: it takes a model
    description, a description of a target programming language and
    set of external particles and generates a sequence of instructions
    that compute the corresponding scattering amplitude as a function
    of external momenta, spins, and other quantum numbers.
    For this purpose, (impure)
    functional programming languages with a strong type system provide
    one of the most convenient environments, and \OCAML/~\cite{O'Caml}
    was selected for
    this task.  As a target programming language, only \FORTRAN/ is
    currently fully supported.  Implementing descriptions of other target
    programming 
    languages is straightforward, however.

  \item The matrix-element code as generated by \OMEGA/ is the time-critical
    part of the 
    program.  It exclusively consists of linear operations applied to
    static objects (four-momenta, spinors, matrices) which are built
    from arrays of complex numbers.  This problem is well suited for
    \FORTRAN/, therefore, \OMEGA/
    produces code in this language.  The interface between \WHIZARD/ and its
    matrix element code is kept strictly and portably \C/-interoperable,
    however, so matrix-element code written in \C/, \Cpp/, or other languages
    with \C/ binding, can easily be substituted.

  \item \VAMP/ -- the oldest part of the package -- is a purely numerical
    library and has therefore also been 
    implemented in \FORTRAN/.  

  \item Third-party libraries accessed by \WHIZARD/ are written in various
    dialects, ranging from \FORTRANSEVENTYSEVEN/ (\CIRCE/) to \Cpp/
    (\HEPMC/).  With the \Cpp/ parts accessed via \C/ interface code,
    the \FORTRAN/ language standard allows \WHIZARD/ to
    interface all of them natively without platform dependencies.

\end{itemize}
While these components are represented as separate libraries in the technical
sense, at the user level \WHIZARD/ acts as a monolithic program that handles
all communication internally.

\subsection{History and New Features}

Work on \WHIZARD/ began in 1998; its original purpose was the
computation of cross sections and distributions for electroweak
processes at a future linear
collider~\cite{Boos:1997gw,Chierici:2001ar}.  In particular, $2\to 6$ 
fermion processes such as vector-boson scattering could not be treated
by the automatic tools available at that time.  The acronym \WHIZARD/
reflects this: W, HIggs, Z, And Respective Decays.  Since then, the
scope of \WHIZARD/ has been extended to cover QCD and hadron collider
physics, the complete SM, and many of its extensions.

Initially, \WHIZARD/ used \MADGRAPH/~\cite{madgraph} and
\COMPHEP/~\cite{comphep} as (exchangeable) matrix-element generators.
Subsequently, these have been replaced as default by the \OMEGA/
optimizing matrix element generator
which avoids the factorial growth of matrix-element
complexity with the number of external particles.  Furthermore,
\WHIZARD/ includes the \VAMP/~\cite{vamp} library for adaptive
multi-channel integration.  In its current state, the \WHIZARD/
project has been merged with the \VAMP/ and \OMEGA/ projects.

For version 2 of the program \WHIZARD/, the program core has been completely
revised with the aim of providing a more conveniently extensible platform that
handles physics processes at hadron colliders in particular.  Amplitudes and
derived quantities are internally represented by a generic \emph{interaction}
structure that describes a correlated quantum state of a set of particles,
which is used throughout the program.  The \WHIZARD/ package as a whole, which
used to consist of several parts connected by Makefiles and scripts, has
become a monolithic program which uses dynamic libraries for extending itself
by compiled matrix-element code at runtime.

On the physics side, the most important addition is support for matrix-element
factorization.  While \WHIZARD/~1 was able to compute complete matrix elements
for multi-particle final states, \WHIZARD/~2 adds the possibility to factorize
processes, e.g., into on-shell production and decay, and thus to handle
situations where complete matrix elements are either computationally
infeasible or undesired for other reasons.  The subprocess factors are
integrated separately and convoluted in the simulation step, retaining color
correlations (in leading-order $1/N_c$), and spin correlations at the quantum
level.  Both exclusive and inclusive particle production can be modeled, as
long as described by leading-order perturbation theory.

\WHIZARD/~2 also simplifies summation over equivalent particles such as quarks
and gluons in the initial state at the LHC.  Parton structure functions are
taken from the LHAPDF library, which is fully supported.  (For
convenience, frequently-used structure functions are also available
for direct access, without installing LHAPDF.)  The energy scale can
be computed event by event using arbitrary kinematic variables.  Running
$\alpha_s$ is available.  Events can be reweighted, read and written in various
recent standard formats (\HEPMC/, \LHEF/).

Another important change is the introduction of a scripting language called
\SINDARIN/ that unifies the tasks of specifying input parameters, declaring
cuts, observables and reweighting factors, and steering a workflow that
includes integration, simulation and analysis, possibly with conditionals and
loops.

\section{Checks and Applications}
\label{sec:appl}


\subsection{Standard Model}
\label{sec:applsm}

\WHIZARD/ supports the complete Standard Model of electroweak and
strong interactions, and reliably computes partonic cross sections for
processes with $4$, $6$, or more particles in the final state, as they
are typical for electroweak high-energy processes such as weak-boson,
Higgs, and top-quark production and decay.  The correctness of the numerical
results, while assured in principle by the validity of the underlying
algorithm, nevertheless should be and has been checked, both by
internal tests and in the context of published physics studies.

\subsubsection{Previous studies}

For instance, in recent work on $W$ pair
production~\cite{Beneke:2007zg}, \WHIZARD/ was used for numerically
computing complete tree-level four-fermion cross sections, in
agreement with analytic calculations.  An exhaustive list of $e^+e^-$
six-fermion cross sections in the SM has been carefully cross-checked
between \WHIZARD/ and \LUSIFER/~\cite{Dittmaier:2002ap}.  All
calculated cross sections were found to agree, taking into account
differences in the treatment of unstable particles that are well
understood.  Six- and eight-fermion final states in top-quark
processes have been studied in Ref.~\cite{Schwinn:2004vd}.

The determination of the Higgs potential will be one of the tasks for
the next generation of colliders.  In a comprehensive study of Higgs
pair production at the LHC and the ILC~\cite{Djouadi:1999gv}, the
analytic results for SM Higgs pair production were numerically
cross-checked with \WHIZARD/, and it could be established that triple
Higgs production in the SM will remain below observability at either
collider.

At SLAC, a large database of SM multi-fermion events in \STDHEP/ format
has been generated using \WHIZARD/~\cite{barklow}, intended for
further experimental and phenomenological ILC studies.  A recent
analysis of possible supersymmetry measurements at the
ILC~\cite{hewett} made use of this database to demonstrate that SM
backgrounds, calculated with complete tree-level matrix elements, are
indeed significantly larger than predicted with the approximations of,
e.g., \PYTHIA/.

In the following, we add to this list a collection of results that apply
specifically to LHC physics.  All results have been obtained using the latest
revision of \WHIZARD/~2.

\subsubsection{W + jets}

The class of processes $W + \mbox{jets}$ at the LHC is interesting by itself,
providing a measurement of partonic luminosity and of $W$-boson properties,
and it constitutes an important background for a plethora of new-physics
processes.  Here, the $W$ actually stands for its decay products $\ell+\nu$,
where $\ell$ may either be an electron or a muon.  In Table~\ref{tab:W+jets},
we list results for $n=2,3,4,5$ jets.  We choose the $e^-\bar\nu_e$ decay of
the $W^-$ for concreteness, and set all fermion masses to zero.  The notation
is: $g$ = gluon, $q$ = quark ($d,u,s,c$), $\bar q$ = antiquark, and $j$ = jet
(gluon, quark, or antiquark).

We specify the following generic cuts for regulating infrared and collinear
singularities (where $j$ denotes a light quark or gluon jet):
\begin{gather}
  \label{eq:cut1}
  p_T(j) > 20 \;\textrm{GeV}
\\
  - 2.5 < \eta(j) < 2.5
\\
  \Delta R(j, j) > 0.7
\end{gather}
as well as the following experimental resolution cuts:
\begin{gather}
    p_T(\ell) > 20 \;\textrm{GeV}
\\
  - 2.5 < \eta(\ell) < 2.5
\\
  \Delta R(j, \ell) > 0.4
  \label{eq:cut2}
\end{gather}
All processes which involve more than one quark-antiquark pair, initial and
final state combined, contain additional photons, $W$ or $Z$ bosons in
intermediate states.  The non-QCD contributions cannot be neglected: while
photon exchange is usually negligible compared to gluon exchange, $W$ and $Z$
bosons can become resonant and effectively lower the order of the process.  On
the other hand, processes with only one quark-antiquark pair are pure QCD
processes with the insertion of one external $W$ boson.

We use the following nonvanishing SM input parameters:
\begin{align*}
  G_F     &= 1.16639\times 10^{-5} \;\textrm{GeV}^{-2}
\\
  m_Z     &= 91.1882\;\textrm{GeV}
\\
  m_W     &= 80.419\;\textrm{GeV}
\\
  \alpha_s &= 0.1178
\\
  \Gamma_Z     &= 2.443\;\textrm{GeV}
\\
  \Gamma_W     &= 2.049\;\textrm{GeV} 
\end{align*}
The value of $\alpha_s$ is kept fixed (it would be possible to vary it
according to some scheme of scale determination).  The CKM matrix is set to
unity.

The $pp$ collider energy is fixed as $\sqrt{s}=14\;\textrm{TeV}$.  We
choose the CTEQ6L set for the proton structure functions.  The sum
over partons is done in the process definition, not by adding
individual subprocesses explicitly.  

The numbers in the tables have been obtained with CKM matrix
set to unity.  Investing some more CPU time, CKM mixing effects can
be incorporated simply by switching the model.

The cross-section results are obtained averaging the integration calls only;
the preceding adaptation calls are for preparing the \VAMP/ integration grids.
The quoted error is the estimate for one standard deviation for the average.
No showering and no matching are applied here. 

\begin{table}[hbt]
  \caption{Results for $W+\textrm{jets}$ processes at the LHC.
    $j=g,d,u,s,c,\bar d, \bar u, \bar s, \bar c$.  All processes are
    computed with the complete Standard Model in all intermediate
    states, CKM matrix set to unity.}
  \label{tab:W+jets}
  
  \begin{equation*}
    \begin{array}{l|cc|cc}
    \hline
    \text{Subprocess} & \text{calls (adapt.)} 
    & \text{calls (integ.)}
    & \sigma\;\textrm{[fb]} & \Delta\sigma [\%]
    \\
    \hline
    gg \to Wq\bar q & 
    \phantom{2}5\,\textrm{M}\phantom{.0} & 
    5\,\textrm{M} & 
    24,091 &
    0.05
    \\
    gg \to Wq\bar qg & 
    \phantom{2}5\,\textrm{M}\phantom{.0} & 
    5\,\textrm{M} & 
    \phantom{2}9,142 &
    0.07
    \\
    gg \to Wq\bar qgg &
    \phantom{2}7.5\,\textrm{M} & 
    5\,\textrm{M} & 
    \phantom{2}2,363 &
    0.3\phantom{0}
    \\
    gg \to Wq\bar qggg & 
    20\,\textrm{M}\phantom{.0} & 
    5\,\textrm{M} & 
    \phantom{22,}524 &
    1.2\phantom{0}
    \\
    \hline
    jj \to Wj & 
    \phantom{1}
    5\,\textrm{M}\phantom{.0}
    &
    5\,\textrm{M}
    & 
    936,230    
    &
    0.03
    \\
    jj \to Wjj & 
    \phantom{1}
    5\,\textrm{M}\phantom{.0}
    & 
    5\,\textrm{M}
    & 
    287,180
    &
    0.05
    \\
    jj \to Wjjj & 
    \phantom{1} 7.5\,\textrm{M}
    & 
    5\,\textrm{M}
    & 
    \phantom{2}
    79,540
    &
    0.08
    \\
    jj \to Wjjjj & 
    10\,\textrm{M}\phantom{.0} & 
    5\,\textrm{M} & 
    \phantom{7}20,900 &
    0.3\phantom{0}
    \\
    \hline
    \end{array}
  \end{equation*}
\end{table}

\subsubsection{Top pairs}

In this section we summarize results for processes that include top pair
production (and Higgs) as intermediate states.  The lepton masses are set
zero.  We use
\begin{align*}
  G_F     &= 1.16639\times 10^{-5} \;\textrm{GeV}^{-2}
\\
  m_Z     &= 91.1882\;\textrm{GeV}
\\
  m_W     &= 80.419\;\textrm{GeV}
\\
  m_b     &= 4.2\;\textrm{GeV}
\\
  m_t     &= 174.0\;\textrm{GeV}
\\
  m_H     &= 120.0\;\textrm{GeV}
\\
  \alpha_s &= 0.1178
\\
  \Gamma_Z     &= 2.443\;\textrm{GeV}
\\
  \Gamma_W     &= 2.049\;\textrm{GeV} 
\\
  \Gamma_t     &= 1.523\;\textrm{GeV} 
\\
  \Gamma_H     &= 0.0036\;\textrm{GeV}
\end{align*}
and the cuts Equs.~(\ref{eq:cut1})-(\ref{eq:cut2}), which are applied
in the same way to $b$ jets as for light quark jets. 
In the processes
with more than six final-state particles, we chose to break down the
sum over flavors into subprocesses: this eliminates a trivial
redundancy that originates from the sum over lepton flavors in
particular, which is not (yet) eliminated by \WHIZARD/.

\begin{table}[hbt]
  \caption{Results for $t\bar t$-related processes at the LHC.
    $j=g,d,u,s,c,\bar d, \bar u, \bar s, \bar c$ and
      $\ell=e,\mu,\tau$}.  All processes are 
    computed with the complete Standard Model in all intermediate
    states, CKM matrix set to unity.
  \label{tab:tt}
  
  \begin{equation*}
    \begin{array}{l|cc|cc}
    \hline
    \text{Subprocess} & \text{calls (adapt.)} & \text{calls (integ.)}
    & \sigma\;\textrm{[fb]} & \Delta\sigma [\%]
    \\
    \hline
    jj \to \ell^+\ell^-\nu\nu b\bar b & 
    10\,\textrm{M}
    & 
    5\,\textrm{M}
    & 
    27,845\phantom{.0}
    &
    0.04
    \\
    jj \to \ell^+\ell^-\nu\nu b\bar b j & 
    10\,\textrm{M} & 
    5\,\textrm{M} & 
    22,780\phantom{.0}
    &
    0.1\phantom{0}
    \\ 
    jj \to \ell^+\ell^-\nu\nu b\bar b j j & 
    10\,\textrm{M} & 
    5\,\textrm{M} & 
    10,500\phantom{.0} &
    0.6\phantom{0}
    \\
    jj \to \ell^+\ell^-\nu\nu b\bar b b\bar b & 
    10\,\textrm{M} & 
    5\,\textrm{M} & 
    \phantom{23,1}73.7 &
    0.9\phantom{0} \\ 
    \hline
    \end{array}
  \end{equation*}

\end{table}


\subsection{Supersymmetry}
\label{sec:applsusy}

Despite its conceptual beauty and simplicity, the minimal
supersymmetric extension of the SM, the MSSM, is a very complicated
model with a large particle 
content and several thousand different vertices. Especially a vast
number of mixing matrices and possible complex phases complicates any
simple relations demanded by the symmetries of the MSSM. A
comprehensive collection of all the Feynman rules of the MSSM and
their derivation can be found in~\cite{jr_kur} and have been
implemented in \WHIZARD/ and \OMEGA/.  The Feynman rules containing
Majorana vertices are implemented according to~\cite{Denner:majorana}.
 
In~\cite{Hagiwara:2005wg}, comprehensive tests have been performed to
verify that the implementation of the MSSM is correct. This has been
done with the help of gauge and supersymmetry invariance checks (Ward
and Slavnov-Taylor identities as described
in~\cite{jr_phd,susyward}). To test all couplings that can be of any
relevance for future experiments in particle physics phenomenology, a
check of more than 500 different processes is necessary. This
extensive list of processes has been tested in~\cite{Hagiwara:2005wg}
by direct comparison with two other public multi-particle event generators,
\SHERPA/~\cite{sherpa} and
\MADGRAPH/~\cite{madgraph,madevent}, showing accordance of all
three programs within the statistical Monte Carlo errors. This
extensive comparison which serves as a standard reference for testing
supersymmetric processes can be found at
\url{http://whizard.event-generator.org}. As a second test, this
implementation has been successfully compared with the MSSM derived
via the \FEYNRULES/ package~\cite{Christensen:2008py}.

With \WHIZARD/ it was for the first time possible to perform
simulations for SUSY processes with six final state particles using
full matrix elements. For the ILC, the
importance of off-shell effects was shown for the production of
e.g.~sbottom pairs in~\cite{Hagiwara:2005wg}.
When using cuts to enhance the signal on top of the
background, interferences of different production amplitudes
(especially heavy Higgses and neutralinos) with identical exclusive
final states lead to deviations from the narrow-width approximation by
an order of magnitude. Similarly, sbottom production at the LHC with
subsequent decay to a $b$ jet and the LSP has been studied
in~\cite{Hagiwara:2005wg}. There, also the contamination of the
tagging decay jets by initial state radiation has been examined,
amounting to the quite complicated process $pp \to \tilde{\chi}_1^0
\tilde{\chi}_1^0 b\bar b b\bar b$ with more than 30,000 Feynman
diagrams, several thousand phase space channels and 22 color flows. It
was shown there, that off-shell effects are important for LHC as well,
and secondly, that there is a severe combinatorial background from ISR
jets.    

Projects that are currently in progress deal with the radiative
production of neutralinos at the ILC~\cite{radiativeneutralino}, the
measurement of SUSY CP phases at the LHC~\cite{lhcsusyphases}, the
determination of chargino and neutralino properties at the
ILC~\cite{KKKRRR} as well as a complete cartography of the edge
structures and spin measurements within the decay chains at the
LHC~\cite{catpiss++}. Further projects deal with the implementation of 
extended GUT- or string-inspired supersymmetric models within
\WHIZARD/ to study their phenomenology~\cite{omwhiz_susyext}.
 
\WHIZARD/ was the first generator for arbitrary multi-leg processes to
contain an implementation of the next-to-minimal supersymmetric SM,
the NMSSM~\cite{NMSSM}. This implementation has been tested in the
MSSM limit as well as by a comparison with an \FEYNRULES/
implementation.


\subsection{Little Higgs}
\label{sec:appllhm}

\WHIZARD/ was the first multi-particle event generator in which
Little Higgs models have been implemented. The Littlest Higgs and
Simplest Little Higgs models are contained in the model library, including
variants of these models like e.g.~different embeddings of the
fermionic content. Several studies for LHC and ILC as well as the
photon collider have been performed with
\WHIZARD/~\cite{omwhiz_little}, focusing especially on the lightest
available states in these models, the so-called pseudoaxions, $\eta$,
pseudoscalar states which are electroweak singlets. The studies are
concerned with production in gluon fusion at the LHC and detection via
rare diphoton decays analogous to the light Higgs, to $t\bar t\eta$
associated production at the ILC, and investigations important for the
model discrimination at the LHC and ILC using special types of
couplings. Ongoing projects deal with general
search strategies at LHC, with focus on the heavy gauge bosons and the
heavy fermionic partners of the quarks in these models. A brief
overview of applications of \WHIZARD/ in the context of Little Higgs
models can also be found in~\cite{appl_reports}.


\subsection{Strongly Interacting Weak Bosons}
\label{sec:applstrong}

If no new particles exist in the energy range of LHC and ILC, and
even the Higgs boson is absent, some scattering amplitudes of
electroweak vector bosons rise with energy and saturate unitarity
bounds in the $\mathrm{TeV}$ range.  This behavior should be
detectable, both by the effects of new strong interactions, possibly
resonances, at the LHC, and by anomalous couplings at lower energies
at both colliders.

Improving on earlier studies~\cite{Boos:1997gw} where due to
computational restrictions, final-state $W$ and $Z$ bosons had to be
taken on-shell, using \WHIZARD/ it became possible to analyze
distributions using complete tree-level matrix elements, vector-boson
decays and all non-resonant background included.  This allowed for
detailed estimates of the ILC sensitivity to those couplings, taking
into account all relevant observables including angular correlations,
without the restrictions of on-shell
approximations~\cite{Chierici:2001ar,Beyer:2006hx}.  Also using
\WHIZARD/ for the simulation, an ongoing ATLAS study will clarify this
issue for the LHC~\cite{kmatrix}. 


\subsection{Exotica}
\label{sec:applexotica}

\WHIZARD/ has also been used to study top quark physics in higgsless
models of electroweak symmetry breaking~\cite{cs_hltop}.

Even the phenomenology of very exotic models can be studied
with~\WHIZARD/, e.\,g.~noncommutative extensions of the
SM~\cite{omwhiz_nc}.  Noncommutative field theories can either be
formulated as nonlocal field theories involving non-polynomial
vertices or as effective field theories with polynomial vertex
factors.  In the first case, it is straightforward to add the
corresponding vertex functions to \OMEGALIB/, while in the latter
case, care must be taken to consistently count the order of effective
vertices.

For a study of exotic models that do not draw enough public attention
to merit a complete supported implementation in \OMEGA/, the easily
readable output of \OMEGA/, allows an alternative approach.  After
generating the SM amplitude with \OMEGA/ in order to
obtain a template with all the required interfaces in place, the
corresponding \FORTRAN/ module can be edited with a text editor,
replacing the SM vertices by the corresponding vertices in
the model and adding diagrams, if required.  The necessary \FORTRAN/ functions
implementing exotic vertex factors can be added to the module, without
having to modify \OMEGALIB/, as discussed in sec.~\ref{sec:hacking}
below. More details about how to add models in general as well as the
interface to the \FEYNRULES/ package~\cite{Christensen:2008py} can be
found in sec.~\ref{sec:buildmodels}.

\section{\OMEGA/: Optimized Matrix Element Generator}
\label{sec:omega}

\OMEGA/~\cite{omega,omega2} is the component of \WHIZARD/ that constructs an optimized
algebraic expression for a given scattering amplitude (or a set of
scattering amplitudes) in a given model.  While it can also be used by
itself as a command line tool (e.\,g.~for producing programs that plot
differential cross sections), it is called by \WHIZARD/ automatically with the
correct arguments when a new process is added to an event generator.

\subsection{Requirements}

For complicated processes, such as searches for new physics at the LHC
and a future ILC, the efficient perturbative computation of scattering
matrix elements has to rely on numerical methods for helicity
amplitudes, at least partially.

The time-honored trace techniques found in textbooks can reasonably
only be used up to~$2\to4$ processes and becomes inefficient for more
particles in the final state.  Therefore, in addition to allowing for
polarized initial and final states, the computation of helicity
amplitudes is the method of choice for complex processes.  While there
are very efficient methods for the analytical calculation of helicity
amplitudes for massless particles, their extension to massive
particles can become cumbersome, while, in contrast, a direct numerical evaluation
results in the most efficient expressions to be used in cross section
calculations and event generation.

It must be stressed that efficiency in the numerical evaluation of
scattering amplitudes is not just a matter of potential wasteful use
of computing resources.  Instead, since the number of required CPU cycles
varies over many orders of magnitude (cf.~fig.~\ref{tab:P(n),F(n)}),
it can be the deciding factor whether simulations are possible at all.

In addition, all realistic quantum field theoretical models of
particle physics employ the gauge principle at some point and are
therefore subject to large numerical cancellations among terms in
perturbative expressions for scattering amplitudes.  Any implementation
that fails to group terms efficiently will suffer from large numerical
uncertainties due to numerically incomplete gauge cancellations.

\OMEGA/ implements an algorithm that collects all common
subexpressions in the sum over Feynman diagrams contributing to a
given scattering amplitude at tree level.  Note that the common
subexpression elimination~(CSE) algorithm in a general purpose
compiler will not be able to find all common subexpressions already in
a moderately sized scattering amplitude, due to the size of the
corresponding numerical expressions.  It remains an open question,
whether amplitudes calculated with twistor-space
methods~\cite{Twistors} could improve on the \OMEGA/ algorithm (for a
comparison that seems quite discouraging for twistor amplitudes,
cf.~\cite{weinzierl}). While the former produce compact analytical
expressions, substantial numerical complexity is hidden in the
effective vertices. Furthermore, the extension to massive particles is
not straightforward~\cite{Schwinn/Weinzierl:Massive-Twistors}.

The building blocks used in \OMEGA/ amplitudes correspond to
expectation values of fields in states of on-shell external particles
\begin{equation}
\label{ref:1POSWF}
  W(x;p_1,p_2,\ldots,p_m) = \braket{p_1,p_2,\ldots,p_n|\phi(x)|p_{n+1},\ldots,p_m}\,.
\end{equation}
In the case of gauge bosons, they satisfy Ward identities, that ensure
that gauge cancellations take place inside these building
blocks~\cite{omega}.

\subsection{Complexity}
\label{sec:complexity}

The irreducible complexity of a given tree level scattering amplitude
is bounded from below by the number of its poles in kinematical
invariants.  In the absence of conserved charges, this number can be
estimated by the number of independent internal momenta that can be
built from the external momenta.  Taking into account momentum
conservation, it is easy to see that there are
\begin{equation}
  P(n) = \frac{2^n-2}{2} - n = 2^{n-1} - n - 1
\end{equation}
independent internal momenta in a $n$-particle scattering amplitude.
This number should be contrasted with the number of Feynman diagrams.
For realistic models of particle physics, no closed expressions can be
derived, but in a one-flavor $\phi^3$-theory, there are exactly
\begin{equation}
  F(n) = (2n-5)!! = (2n-5)\cdot(2n-7)\cdot\ldots\cdot3\cdot1
\end{equation}
tree level Feynman diagrams contributing to a $n$-particle scattering
amplitude.

\begin{table}
  \begin{center}
     \begin{tabular}{r|r|r}
       $n$ & $P(n)$& $F(n)$ \\\hline
         4 &     3 & 3      \\
         5 &    10 & 15     \\
         6 &    25 & 105    \\
         7 &    56 & 945    \\
         8 &   119 & 10395  \\
        10 &   501 & 2027025 \\
        12 &  2035 & 654729075 \\
        14 &  8177 & 316234143225 \\
        16 & 32751 & 213458046676875
     \end{tabular}
  \end{center}
  \caption{\label{tab:P(n),F(n)}
    The number of $\phi^3$ Feynman diagrams~$F(n)$ and independent
    poles~$P(n)$.}
\end{table}
Obviously, $F(n)$ grows much faster with~$n$ than~$P(n)$
(cf.~table~\ref{tab:P(n),F(n)}) and already for a~$2\to6$ process, we
find that the number of poles is two orders of magnitude smaller than
the number of Feynman diagrams.

For realistic models with higher order vertices and selection rules at
the vertices, empirical evidence suggests
\begin{equation}
  P^*(n) \propto 10^{n/2}
\end{equation}
while the factorial growth of the number of Feynman diagrams remains
unchecked, of course.

While~$P(n)$ is a priori a lower bound on the complexity, it turns out
that this bound can approached in numerical~\cite{alpha,helac} and
symbolic~\cite{omega,omega2} implementations.  Indeed, the number of
independent momenta in an amplitude is a better measure for the
complexity of the amplitude than the number of Feynman diagrams, since
there can be substantial cancellations among the latter.  Therefore it
is possible to express the scattering amplitude much more compactly than by
a sum over Feynman diagrams.

\subsection{Relations to other algorithms}
\label{sec:relations}

Some of the ideas that \OMEGA/ is based on can be traced back to
\HELAS/~\cite{HELAS}.  \HELAS/ builds Feynman amplitudes by recursively
forming off-shell `wave functions'~(\ref{ref:1POSWF})
from joining external lines with
other external lines or off-shell `wave functions'~(\ref{ref:1POSWF}).

The program \MADGRAPH/~\cite{madgraph} automatically generates
Feynman diagrams and writes a \FORTRAN/ program corresponding to their
sum.  The amplitudes are calculated by calls to \HELAS/.
\MADGRAPH/ uses one straightforward optimization: no statement is
written more than once.  Since each statement corresponds to a
collection of trees, this optimization is effective for a moderate number
(say, four) of particles in the final state.  Since the amplitudes are
given as a sum of Feynman diagrams, this optimization does not remove
the factorial growth of the computational complexity with the
  number of external particles.  

The symbolic algorithm of \OMEGA/ which is analogous in
  structure to the numerical algorithms of \ALPHA/~\cite{alpha} and 
\HELAC/~\cite{helac}, ensures, by design, that for any given helicity
amplitude no independent subexpression is evaluated twice.  It thus
allows for the optimal asymptotic behavior.  In practice, further
optimizations are a matter of the concrete implementation, programming
language, and compiler.

\subsection{Architecture}
\label{sec:architecture}

\OMEGA/ does not follow the purely numerical approach
of~\cite{alpha,helac}, but constructs a symbolic representation of an
optimally factored scattering amplitude instead, that is later
translated to a programming language (\FORTRAN/) for
compilation into machine code by a corresponding compiler.  The
symbolic approach brings three advantages:
\begin{enumerate}
   \item in principle, 
     the compilation to machine code allows a faster execution speed
     than the numerical programs that have to loop over large arrays.
     In practice this allows a two- to four-fold increase in execution
     speed, depending on the process and the model under consideration.
   \item the intermediate \FORTRAN/ code is human-readable
     (cf.~Appendix~\ref{sec:sample-code})
     and can easily be edited in order to experiment with the
     implementation of very exotic models, radiative corrections or
     form factors (cf.~section~\ref{sec:hacking}).
   \item more than one programming language for the numerical
     evaluation can be supported.
\end{enumerate}
For our applications, the second advantage is particularly important.

Since it is an exclusively symbolic program, \OMEGA/ has been
implemented in the impure functional programming language
\OCAML/~\cite{O'Caml}.  \OMEGA/ makes extensive use of the advanced
module system of \OCAML/ and is structured in small modules
implementing abstract data types with well defined interfaces and
strong type-checking.

The crucial feature of \OMEGA/ is that it internally represents the
scattering matrix element not as a tree of algebraic expressions, but
as a Directed Acyclical Graph~(DAG), where each subexpression appears
only once, instead.  In principle, it would be possible to construct
first the tree corresponding to the sum of Feynman diagrams and to
transform it subsequently to the equivalent minimal DAG by an
algorithm known as Common Subexpression Elimination~(CSE) in
optimizing compilers.  However, the size of the expressions resulting
from the combinatorial growth with the number of external particles
makes this all but impossible for practical purposes.

The basic algorithm of \OMEGA/ proceeds as follows:
\begin{enumerate}
  \item in a first step, functions from the \texttt{Topology} module
    construct the DAG corresponding to the sum over all Feynman
    diagrams in unflavored $\phi^n$-theory, where~$n$ is the maximal
    degree of the vertices in the model under consideration.  The
    abstract DAG data type is implemented by the \texttt{DAG} functor
    applied to a module describing the maximum degree of the vertices
    in the Feynman rules of the model.  It should be stressed that the
    algorithms in \OMEGA/ place no limit on~$n$ and are therefore
    applicable to arbitrary effective theories and can support
    skeleton expansions of radiative corrections.
  \item in a second step, all possible flavorings of the DAG are
    derived from the Feynman rules of the model encoded in a module
    implementing the \texttt{Model} signature.  The algorithm ensures
    the symmetry and antisymmetry of the scattering amplitude for
    identical bosons and fermions.  In the case of Majorana fermions,
    the standard algorithm for Feynman diagrams~\cite{Denner:majorana}
    has been implemented for DAGs as well~\cite{jr_phd}.  Together
    with the numerical expression for each vertex and external state,
    this flavored DAG is the minimal faithful representation of the
    scattering amplitude.

    During this step, it is possible to select
    subamplitudes, e.\,g.~by demanding that only contributions
    containing a certain set of poles are included.  While this
    restriction to cascade decays can break gauge invariance because
    it doesn't select gauge invariant subsets~\cite{groves}, it is
    nevertheless a useful feature for testing the accuracy of commonly
    used approximations.
  \item finally, a module implementing the \texttt{Target} signature
    is used to to convert the internal representation of the DAG (or
    of a set of DAGs) into code in a high level programming language
    (currently only \FORTRAN/ is fully supported), that will
    be compiled and linked with the rest of \WHIZARD/.
\end{enumerate}

This modular structure is supported by a library of purely functional
abstract data types that implement, among others, combinatorial
algorithms (\texttt{Combinatorics}), efficient operations on tensor
products (\texttt{Products}) and the projective algebra of internal
momenta subject to overall momentum conservation (\texttt{Momentum}).

The implementation of models as an abstract data type allows \OMEGA/
to apply functors to a model and derive related models.  In fact, such
a functor (\texttt{Colorize}) is used internally to add color in the
color-flow basis (cf.~section~\ref{sec:color}) to models from the
$SU_C(N)$ representations~\cite{omega2}. As another example, it is
also possible to automatically derive the $R_\xi$-gauge version of a
spontaneously broken gauge theory from the Feynman rules in unitarity
gauge~\cite{cs_phd}.  Parsers for external model description files can
also be implemented as a special case of \texttt{Model}.

While it is in principle possible to treat color as any other quantum
number and to generate the amplitudes for each possible color flow
independently (a fact that was used in version~1 of \WHIZARD/), it is
much more efficient and convenient to add color internally with the
\texttt{Colorize} functor and to generate the amplitudes for all color
flows at once~\cite{omega2}.  While this approach uses exactly the
same algorithm as before, the added infrastructure for multiple
amplitudes allows for additional eliminations of common subexpressions
amongst amplitudes with different colors and flavors in the external
states.  This new approach resulted in an orders-of-magnitude speedup
in the generation of colored matrix elements from version~1 to
version~2 of \WHIZARD/.

\section{The \WHIZARD/ Architecture}

To promote a matrix-element generator such as \OMEGA/ to a complete
automated physics simulation program, it has to be supplemented by a
variety of program elements.  We have implemented all necessary parts
in the program package \WHIZARD/.  (This program name also stands for
the event generator as a whole, including the matrix-element
generating part.)

First of all, \WHIZARD/ provides code that accounts for suitable
phase-space parameterizations (phase-space channels), that selects
among the multitude of possible channels, that integrates phase space
with generic cuts in order to compute inclusive quantities, and that
samples phase space in order to generate a sequence of exclusive (in
particular, unweighted) scattering events.  Since the computing cost
of phase-space sampling, if done naively, can easily exceed the
computing cost of optimized matrix-element computation by orders of magnitude,
the algorithms must be able to keep this cost down at a manageable
level.

Furthermore, \WHIZARD/ implements (or contains interfaces to) the
physics models, to beam dynamics such as parton structure,
initial-state radiation, beamstrahlung, beam polarization etc., to
final-state showering and hadronization, and to external event-file
processing, detector simulation, and analysis.

The top-level routines that combine these parts to a working event generator,
depending on a user-provided script that contains all necessary input, are
also part of \WHIZARD/.  Finally, \WHIZARD/ contains a lightweight analysis
module that can plot distributions of various physical observables without the
need for external programs.

\subsection{Core Libraries}

Viewed from the user perspective, \WHIZARD/ behaves as a monolithic program.
In technical terms, it consists of a core library, which in turn is
broken down to modules (in the \FORTRAN/ sense), several additional
libraries (some of them third-party) which are linked at program startup, and
one or more process-code libraries that are typically generated and
dynamically linked at runtime.  The library structure is reflected in the
directory structure of the \WHIZARD/ installation; each subpackage is located
in a separate subdirectory of the \texttt{src} subdirectory.

The \WHIZARD/ program is configured and built using \texttt{autotools}; in
particular, the build process of the program itself is managed by
\texttt{libtool}.  Therefore, both statically and dynamically linked versions
of the libraries and the executable can be built, controlled by the options to
\texttt{configure} that the administrator specifies during installation.  In
addition, the \SINDARIN/ language supports, at runtime, the building of a
statically linked executable that includes specific process code and can be
launched separately.

In the following, we describe the individual libraries that constitute
\WHIZARD/.  The \FORTRAN/ part is organized into modules.  They roughly
resemble \Cpp/ class implementations: typically, they define a public data
type together with the methods that operate on it.  Among the methods are
initializers and finalizers that are used to dynamically create and destroy
objects of the data type.  Most data types are opaque, so access to their
internals is confined to the respective module.  Some modules contain several
data types that access each other's internals, and some contain parameter
definitions, static data, or additional functions or subroutines that are not
classified as methods.  Note that \FORTRAN/, unlike \Cpp/, has no header files
but enforces a module dependency hierarchy, which is automatically cared for
by the Makefile setup.

\subsubsection{\texttt{libaux}}

  The auxiliary library \texttt{libaux} contains basic modules which are
  required by more than one library.  They include \texttt{kinds} (real and
  integer kinds, matching \FORTRAN/ and \C/ definitions),
  \texttt{iso\_varying\_string}, the \texttt{limits} module that exhibits all
  fixed parameters accessed by the core library, modules that interface the
  operating system and implement the paths and parameters set by
  \texttt{configure} , the \texttt{diagnostics} module that centralizes error
  handling, and basic physics modules which define constant and data types and
  functions for three- and four-vectors and Lorentz transformations, used
  throughout the program.

\subsubsection{\texttt{libwhizard\_core}}

  The modules that make up the core library \texttt{libwhizard\_core} are
  loosely organized as module groups:
  \begin{itemize}
  \item 
    Auxiliary modules: OS interface, CPU timing, hashing, sorting, and
    permutations, string formatting, and the MD5 algorithm.
    The latter is used for checking whether certain data have to be recreated
    or can be reused.
  \item
    Text-handling modules: An internal-file implementation based on ISO
    varying strings, a generic lexer, syntax table handling, and a
    generic parser.  Using these modules, \WHIZARD/ is able to
    take a syntax description and translate textual input into an internal
    parse-tree representation.
  \item
    Tools for physics analysis: Histograms and such, the data types for
    particle aliases and subevents, the implementation of variable lists, and
    the \SINDARIN/ expression handler.  The latter implements methods to
    compile a parse 
    tree into an evaluation tree and to evaluate this compiled version.
  \item
    The module that manages physics models: parameters, particle data, and
    vertices as used for phase-space construction.
  \item
    Quantum number definition and implementation: Helicity, color, and flavor.
  \item
    Correlated quantum numbers: State (density) matrices, interactions, and
    evaluators for matrix elements.
  \item
    Particle data types and objects, including event formats.
  \item
    Beam definition.
  \item
    Beam spectra and structure functions.  This includes beamstrahlung
    (\CIRCE/ interface), ISR, EPA, and the \LHAPDF/ interface for parton
    structure functions.
  \item
    Phase space: Mappings of kinematics to the unit hypercube, the phase-space
    tree structure, phase-space forests that implement a multi-channel
    structure, and the \texttt{cascades} module that constructs phase-space
    channels for a given process.
  \item
    Process code interface: a module that handles process
    libraries, the \texttt{hard\_interaction} module that interfaces the
    matrix element, and the \texttt{processes} module that combines the hard
    interaction with its environment and interfaces the \VAMP/ integration
    algorithm.  Furthermore, there is a \texttt{decays} module that implements
    decay chains, and the \texttt{events} module that collects processes and
    decays and provides event I/O.
  \item
    The \texttt{slha\_interface} module provides the reader for SUSY
    parameter input files that follow the SUSY Les Houches Accord format.
  \item
    A set of modules that implement the high-level methods for runtime data,
    compilation, integration, and simulation.
  \item
    The \texttt{commands} module compiles and executes
    the command part of the \SINDARIN/ language.
  \item
    The \texttt{whizard} module implements the global initializer and finalizer
    and various methods for accessing \WHIZARD/.
  \item
    The \texttt{main} program realizes \WHIZARD/ as a stand-alone program,
    including the interpreter for command-line options.
  \end{itemize}

\subsubsection{\texttt{libvamp}}

The \VAMP/ integration library is self-contained, while it smoothly integrates
into the \WHIZARD/ setup.  It contains no reference to the specific physics of
a Monte-Carlo generator for scattering processes (for instance, it has no
notion of four-vectors), but rather implements a multi-channel version of the
well-known \VEGAS/ algorithm that can be used for integrating arbitrary
functions of real variables over the multi-dimensional hypercube.  In addition
to the Monte-Carlo integration mode, it also provides an event-generation mode
with simple rejection to generate unweighted events.

\VAMP/ is a stand-alone package with its own \texttt{configure} and
\texttt{make} process.  In the context of \WHIZARD/, these tasks are accounted
for by the main \texttt{configure} and \texttt{make}, and the library is
automatically built and linked into \WHIZARD/.

\subsubsection{\texttt{libomega} and \texttt{libmodels}}

Similar to \VAMP/, the \OMEGA/ package is self-contained.  In the \WHIZARD/
context, the \OMEGA/ build process is automatically triggered by the main
\texttt{configure} and \texttt{make} process.

The matrix-element generator \OMEGA/ is not linked to the \WHIZARD/
executable.  Instead, it exists as a stand-alone program in the library area
of the \WHIZARD/ installation (more precisely, there is a separate program for
each model supported by \OMEGA/), and \WHIZARD/ calls this program via the
operating system when it is needed.  \WHIZARD/ also arranges for the resulting
code being compiled, organized into process libraries with their interface
code, and dynamically linked at runtime.

To enable dynamic linking in a portable way, the matrix-element interface is
strictly \C/ interoperable.  This has the side effect that, by manual
intervention, matrix element code written in \C/ or some other \C/-compatible
language can replace the automatically generated code, as long as the
interface is implemented completely.

For the matrix-element code to be executable, \WHIZARD/ must also link the
\OMEGA/ runtime library \texttt{libomega}.  This library, which is written in
\FORTRAN/, contains the data type definitions used in the matrix
element code: vectors, spinors, matrices.

The current \OMEGA/ version requires an additional library \texttt{libmodels}
which deals with the plethora of model-specific parameters that occur in the
various BSM scenarios.  The corresponding source code is located in
\texttt{src/models}.

\subsubsection{\texttt{libcirce1} and \texttt{libcirce2}}

The \CIRCE/1 library implements convenient parameterizations that
describe beamstrahlung, i.e., the macroscopic emission of photons from
colliding $e^\pm e^-$ beams.  Obviously, it is useful only for this
specific collider setup, and the parameterizations have been obtained
for a finite set of collider parameters only.  Therefore, one of these
parameter sets (collider type and energy) must be specified as-is when
\CIRCE/1 is to be used. 

\CIRCE/1 and \CIRCE/2 are both written in \FORTRANSEVENTYSEVEN/ which
is completely compatible with the current \FORTRAN/ standard, thus
they are directly interfaced by \WHIZARD/. The \CIRCE/2 library is
actually an implementation which describes $e\gamma$ and
$\gamma\gamma$ collisions.

\subsection{Optional (Third-Party) Libraries}

Not all tasks necessary for event generation are handled natively by \WHIZARD/
and its components \VAMP/ and \OMEGA/.  For some tasks, \WHIZARD/ merely acts
as a broker for libraries that exist independently and may or may not be
present in the installation.

\subsubsection{\LHAPDF/}

\WHIZARD/ provides its own set of standard parton distribution
functions.  If other parton distribuction functions are needed, they
can be accessed via the \LHAPDF/ library. In that case, the
\LHAPDF/ library must be installed on the
system, it is not part of the \WHIZARD/ installation.  The \texttt{src/lhapdf}
subdirectory merely provides a non-functional replacement library that is
linked when \LHAPDF/ is not available.

All structure functions known to \LHAPDF/ (and installed as data files) are
available to \WHIZARD/.  Furthermore, \WHIZARD/ can access the $\alpha_s$
function of the \LHAPDF/ package, so an exact match between the structure
function and the hard process is possible.  Note that \WHIZARD/ is a
leading-order program, therefore the NLO structure functions provided by
\LHAPDF/ have limited applicability.

\subsubsection{\STDHEP/}

\WHIZARD/ supports the \STDHEP/ binary event-file format for output, if the
\STDHEP/ library is linked.  Analogous to \LHAPDF/, this must be installed
separately on the system.  The \texttt{src/stdhep} subdirectory contains a
non-functional replacement if the library is not available.

\subsubsection{\HEPMC/}

\HEPMC/ is a \Cpp/ class library intended for storing, retrieving,
manipulating, and analyzing high-energy physics events.  \WHIZARD/ makes use
of this library for reading and writing event files in a machine-independent
ASCII format.  To make this possible, it contains a portable \Cpp/ wrapper for
\HEPMC/, since the class library itself does not provide a portable interface
at the operating-system level.  The wrapper library accesses the \Cpp/ objects
via pointers and matches each method by a corresponding \texttt{extern "C"}
function.

\WHIZARD/ accesses the \HEPMC/ (wrapper) by corresponding
\texttt{type(c\_ptr)} objects and methods on the \FORTRAN/ side, so it employs
an object-oriented portable \FORTRAN/ API for all functionality that it needs.
In particular, the ASCII event files are not read or written directly, but
only by creating \HEPMC/ event objects and calling the appropriate I/O
methods.

\HEPMC/ is an external library, and its features are accessible only if
\HEPMC/ is properly installed on the system.  If not, a non-functional dummy
library is substituted.

We have chosen \HEPMC/ as a default machine-independent format since it allows
to store essentially all information that is present in \WHIZARD/ events, once
quantum-number correlations are eliminated.  In particular, \HEPMC/ supports
color, and it provides containers for event weights.  There is a limitation,
however: currently, the dedicated \HEPMC/ support for polarization is
insufficient for generic particles, therefore polarization is not fully
supported by this event format.  (In principle, it would be possible to abuse
the \HEPMC/ weight containers for storing polarization information and
correlated quantum numbers, but this is cumbersome and non-standard.)

As an alternative, the \LHEF/ (Les Houches Event Format) is supported,
currently for output only.  This ASCII format is a standard for communicating
with parton-shower programs, otherwise it has more limitations than \HEPMC/.

For internal use, \WHIZARD/ writes and reads event data in a binary format,
using unformatted \FORTRAN/ I/O.  This format contains the complete event
information in a compact form, but it is portable only between machines with
similar architecture and, presumably, the same \FORTRAN/ compiler.

\subsection{Further Components of the Package}

\subsubsection{\texttt{gamelan}}

For graphical visualization, \WHIZARD/ employs the \METAPOST/ program which
is part of the \TeX\ family.  The program takes a graph
description and translates it into native \POSTSCRIPT/ code.  Being integrated
in the \TeX\ system, it has the full power of \TeX\ (and \LaTeX) at hand for
typesetting labels and other textual parts of a graph.

A shortcoming of \METAPOST/ is the lack of native support for floating-point
numbers.  As a workaround, the original \METAPOST/ system contains the
\texttt{graph.mp} package that emulates floating-point numbers for data
plotting.   The macro package \GAMELAN/ builds upon and greatly expands
\texttt{graph.mp}, aimed at flexible and convenient visualization of
two-dimensional data, as it is abundant in high-energy physics.

The complete \GAMELAN/ system resides in a subdirectory \texttt{src/gamelan}.
In the context of \WHIZARD/, the program is called upon request to transform
histogram and other tabulated data from files into \POSTSCRIPT/ plots.  Only a
few features of \GAMELAN/ are actually accessible by the user interface in
\SINDARIN/ code, just sufficient to provide default parameters and drawing
modes.  However, it is always possible to modify and improve the generated
plots by manually editing the automatically generated files.  These are
actually \LaTeX\ files (extension \texttt{.tex}) with embedded \GAMELAN/ code.

\subsubsection{\texttt{feynmf}}

The \FEYNMF/ package is similar, but not identical, to \GAMELAN/.  It
implements Feynman-graph drawing and also generates \POSTSCRIPT/ files.  The
package is independent of \WHIZARD/ and included for convenience.  \WHIZARD/
uses it to visualize the tree structure of phase-space channels.

\section{Algorithms}

\subsection{QCD and Color}
\label{sec:color}

While, in principle, the helicity state of a particle can be
experimentally resolved, the color state of a parton is unobservable.
Nevertheless, models for showering and hadronizing colored partons
require that the program keeps track of color connections among
final-state and, in the case of hadronic collisions, initial-state
particles.  This implies that much of the efficient color-algebra
calculation methods commonly used for the analytical calculation of
QCD amplitudes are not transferable to exclusive event generation,
since the color connection information needed for parton-shower models
is lost.

This fact is well known, and color connections are supported by
Monte-Carlo generators such as \PYTHIA/ and \HERWIG/.  Since
these programs, in most cases, construct amplitudes by combining $2\to
2$ scattering processes with decay and showering cascades, keeping
track of color is essentially straightforward.  The problem becomes
much more involved if many partons take part in the initial
scattering, as it is the case for typical \WHIZARD/ applications.

A possible implementation of color, realized in \MADGRAPH/, makes color
connections explicit: The amplitude is distributed among the possible color
diagrams, which can be either squared including interferences (for the matrix
element squared), or dropping interferences (for the parton shower model).

A purely numerical approach could also treat color as a random
variable which is sampled for each event separately, or summed over
explicitly.  However, the individual colors of partons (e.g., ``red'',
``green'', ``blue'') are not directly related to the color connections
used in parton shower and hadronization models, therefore this
approach is not useful without modification.  Sampling color
connections as a whole would be possible; summing over them for each
event leads back to the \MADGRAPH/ algorithm.

The \WHIZARD/ and \OMEGA/ treatment of color takes a different road.  It relies on
the observation that the $SU(3)$ algebra of QCD is formally equivalent to a
$U(3)=SU(3)\times U(1)$ algebra with a $U(1)$ factor
subtracted~\cite{Maltoni:2002mq}.  In terms of Feynman graphs, in unitarity
gauge, each $SU(3)$ gluon is replaced by a $U(3)$ gluon and a $U(1)'$
ghost-gluon (gluon propagator with negative residue).  The $U(1)'$ gluon does
not couple to three- and four-gluon vertices, so we have to include ghost
diagrams just for gluon exchange between quarks.  Ghost gluons are also kept
as external particles, again with a minus sign, where they cancel the
contributions of external $U(1)$ gluons.  (Note that these ghosts are not to
be confused with the Fadeev-Popov ghosts in a generic gauge.)

This idea leads to a very simple algorithm with the advantage that the
color algebra becomes trivial, i.e., all color factors are powers
of~$3$.  Color connections are derived by simply dropping the ghost
diagrams.  Nevertheless, the squared matrix element computed by adding
gluon and ghost diagrams exactly coincides with the matrix element
computed in the $SU(3)$ theory.

\subsection{Phase Space and Performance}

For computing inclusive observables like a total cross section, the
exclusive matrix element (squared) returned by, e.g., \OMEGA/, has to
be integrated over all final-state momenta.  If the initial beams have
nontrivial structure, there are also integrations over the
initial-parton distributions.

The integration dimension for typical problems (e.g., $2\to 6$
scattering) exceeds 10, sometimes 20.  Only Monte-Carlo techniques are
able to reliably compute such high-dimensional integrals.  This has
the advantage that the algorithm can be used not just for integration,
but also for simulation: a physical sequence of scattering or decay events
also follows a probabilistic law.  It has the disadvantage that the
integration error scales with $c/\sqrt{N}$, where $c$ is a constant,
and $N$ the number of events.  Since $N$ is practically limited even
for the fastest computers (a factor $10$ improvement requires a factor
$100$ in CPU time), all efforts go into minimizing $c$.

For a uniform generation of physical events, in theory one should take
a mapping of phase space that, via its Jacobian, transforms the
kinematic dependence of the actual matrix element with its sharp peaks
and edges into a constant defined on a hypercube.  Such a mapping
would also make integration trivial.  Needless to say, it is known
only for very special cases.

One such case is a scattering process defined by a single Feynman
diagram of $s$-channel type, i.e., the initial partons fuse into a
virtual particle which then subsequently decays.  Here, phase space
can be factorized along the matrix element structure in angles and
invariant-mass variables, such that up to polynomial angular
dependence, each integration depends on only one variable.  For this
case, we need mapping functions that transform a power law (massless
propagator) or a Lorentzian (massive propagator) into a constant.
Mapping polynomials is not that important; the angular dependence is
usually not analyzed in detail and taken care of by rejection methods.
The other two mappings provide the basis for constructing phase space
channels in many algorithms, including the one of \WHIZARD/.

This simple construction fails in the case of $t$-channel graphs,
where massless or massive particles are exchanged between the two
initial particles.  The overall energy constraint does not correspond
to a line in the Feynman graph.  However, the dependence of the
exchanged propagator on the cosine of the scattering angle is still a
simple function.  Therefore, for finding a suitable parameterization,
we flip the $t$-channel graph into a corresponding $s$-channel graph,
where the $z$-axis of integration is aligned to one of the initial
partons, so this polar angle becomes an integration variable.  

Multiple exchange is treated by repeated application of this
procedure.  Flipping graphs is not unique, but for any choice it
reduces $t$-channel graphs to $s$-channel graphs which typically also
exist in the complete matrix element.

The main difficulty of phase-space sampling lies in the simultaneous
presence of many, sometimes thousands, of such graphs in the complete
matrix element.  Neglecting interferences, one can attempt a
multi-channel integration where each parameterization is associated
with a weight which is iteratively adapted until it corresponds to the
actual weight of a squared graph in the integral.  Since there are as many phase
space channels as there are Feynman graphs, without further
optimization the computing cost of phase space scales with the
number of graphs, however.

Since the \OMEGA/ algorithm results in a computing cost scaling better
than the number of graphs, computing \WHIZARD/ phase space should also
scale better, if possible.  To our knowledge, for phase-space
integration there is no analog of the \OMEGA/ algorithm  that accounts
for interference.  Hence, \WHIZARD/ uses heuristics to keep just the
most important phase space channels and to drop those that would not
improve the accuracy of integration.  To this end, it constructs
Feynman graphs for the process, keeping track of the number of
resonances or massless branchings, and dropping terms that fail to meet
certain criteria.  The remaining number of phase space channels (which
might come out between a few and several thousand) is then used as the
basis for the \VAMP/ algorithm which further improves the mappings
(see below).  After each \VAMP/ iteration, the contributions of all
channels are analyzed, and unimportant channels are dropped.

While this is not a completely deterministic procedure, with slight
improvements and tunings it has turned out to be stable and to cover
all practical applications.  By construction, it performs well for
``signal-like'' processes where multiply-resonant Feynman graphs give
the dominant contribution to the matrix element, and subdominant
graphs are suppressed.  In contrast to the \PYTHIA/ approach which
considers only the resonant graphs in the matrix element,
\WHIZARD/ does include all Feynman graphs in the matrix elements and
returns the complete result.  Only the method of integration takes
advantage of the fact that dominant graphs dominate phase space.

``Background-like'' processes like multiple QCD parton production
without resonances, at first glance, appear to be not covered so well
since the number of dominant graphs is not restricted and becomes
large very quickly.  This case has not been tested to the extreme with
\WHIZARD/, although for $2\to 6$ QCD processes it still gives stable
results.  However, fixed-order perturbation theory is not viable for a
large number of partons (unless cuts are very strict, such that the
cross section itself becomes unobservable), and parton-shower methods
are suited better.  With the caveat that proper matching of matrix
element and parton shower is not yet implemented for the CKKW(-L)
algorithm (while MLM matching exists as a separate module), we can
conclude that the \WHIZARD/ phase-space algorithm covers all cases
where the fixed-order matrix element approximation is valid.

For a Monte-Carlo cross section result, the decisive performance
criterion is the value of $c$ in $\Delta\sigma/\sigma = c/\sqrt{N}$.
After adaptation, in typical applications such as electroweak $2\to 6$
processes, a \WHIZARD/ run typically returns a number of order~$1$, so
with $10^6$ events a relative error in the per-mil range can be
expected.  In simple cases the accuracy can become much better, while the
performance will be worse if phase space is not that well behaved, such as in
pure QCD processes.

Another important criterion for a Monte-Carlo algorithm is its ability
to identify the maximum weight of all events, and the fraction of this
maximum that an average event gives.  This determines the reweighting
efficiency for generating unweighted event samples, and if many events
are required, the overall computing cost drastically depends on this
efficiency.

\WHIZARD/ keeps track of the reweighting efficiency.  With \WHIZARD/'s
selection of phase space channels and \VAMP/'s adaptive sampling, in
applications with multiple partons and $t$-channel graphs it typically
ends up in the per-mil- to percent range, while in favorable cases
(multiply resonant, i.e., signal-like), efficiencies of order $10\,\%$
are common.  Given the fact that for a meaningful cross section
result, the number of events in the integration step is often a factor
100 higher than the number of unweighted events needed in the
subsequent simulation, with efficiencies in this range the computing
cost of adaptation, integration, and event generation averages out.

\subsection{Multi-Channel Adaptive Sampling: \VAMP/}

For multi-dimensional integration, \WHIZARD/ makes use of the \VAMP/
integration package~\cite{vamp}.  The \VAMP/ algorithm is an extension
of the \VEGAS/ algorithm~\cite{Lepage:1980dq}.  The \VEGAS/ algorithm
introduces a multi-dimensional rectangular grid in the integration
region.  For each iteration, a given number of events (e.g., $10^6$)
is distributed among the cells, either on a completely random basis
(importance sampling) or evenly distributed among the grid cells, but
randomly within each cell (stratified sampling).  For stratified
sampling, usually the number of cells of the original grid (e.g.,
$20^{15}$) is too large for filling each of them with events, so an
auxiliary super-grid with less cells is superimposed for this purpose
(pseudo-stratification), and within each super-cell, the events
randomly end up in the original cells.

After each integration pass, the sum of integrand values, properly
normalized, yields an estimator for the integral, and the binning of
each dimension is adapted to the shape of the integrand.  For
importance sampling, the adaptation criterion is the integral within each bin,
while for stratified sampling, the bins are adapted to the variance
within each bin.  In practice, for high-dimensional Feynman integrals
both importance sampling and stratified sampling give results of
similar quality.

The \VAMP/ algorithm~\cite{vamp} combines this method with the technique of
multi-channel sampling~\cite{Kleiss:1994qy}.  All selected phase-space
parameterizations, properly mapped to the unit hypercube, are sampled
at once, each one with its own \VEGAS/ grid.  The estimator of the
integral is given by the weighted sum of the individual estimators,
where the weights $\alpha_i$ are initially equal (with
$\sum\alpha_i=1$), but are also adapted after each iteration,
according to the channel-specific variance computed for this
iteration.

The \VEGAS/ algorithm has the effect that peaks in the integrand,
during the adaptation process, become flattened out because after
adaptation more events are sampled in the peak region than elsewhere.
This works only for peaks that are aligned to the coordinate axes.
Using \VAMP/, one tries to arrange the parameterizations (channels)
such that each peak is aligned to axes in at least one channel.  Since
the integrand in any channel is corrected by the Jacobian of the
transformation to the other channels (see Ref.~\cite{vamp} for
details), in effect peaks are removed from all channels where they are
not aligned, and flattened out in those channels where they are.  As a
result, after adaptation, within each channel the effective integrand
is well-behaved, and both the integration error and the reweighting
efficiency are significantly improved.

This adaptation proceeds on a statistical basis, and for reasonable
numbers of events and iterations it is a priori not guaranteed that an optimum
is reached.  In particular, fluctuations become overwhelming when the
number of channels, i.e., degrees of freedom, becomes too large.
However, with the selection of phase-space parameterizations done by
\WHIZARD/, the algorithm has proved sufficiently robust, such that it is
universally applicable to the physics processes that \WHIZARD/ has to
cover.

\subsection{Interactions and Evaluators}

The possible states of a quantum systems can be described by a generic density
matrix, which is differential in all quantum numbers of the objects it
involves.  In a specific basis, the density matrix normally incorporates both
diagonal and non-diagonal elements.  The latter are usually referred to as
quantum correlations or entanglement.  If a system is composed of multiple
elementary objects, its density matrix may or may not factorize into a product
of individual density matrices.  If it is diagonal but non-factorizable, the
state is classically correlated.  If it factorizes, it is uncorrelated.

\subsubsection{State Matrices}

Physical events, described as particles after interacting with a detector, are
uncorrelated by definition.  However, at intermediate stages of a high-energy
physics calculation, correlated states have to be described.  This is
reflected by the internal representation in the \WHIZARD/ code.

The representation makes use of the fact that, if a state is generated by a
Monte-Carlo integration algorithm it has a well-defined momentum for all of
its particles.  In other words, the kinematical variables can be treated as
classical and uncorrelated.

The quantum numbers for which correlation have to be implemented are flavor,
color, and helicity.  A quantum state is therefore represented in \WHIZARD/ by
a \texttt{state\_matrix} object, which is a list of allowed quantum-number
combinations for a system of $n$ particles together with their associated
amplitudes, complex numbers.  (The internal representation is actually a
tree.)  The same representation, with an adequate interpretation of the
entries, is used for describing squared amplitudes or interference terms.

Flavor correlations can be treated as classical for the purposes of
Monte-Carlo simulation.  Thus, a state matrix has one flavor entry per
particle (which may be undefined flavor).

As explained above (Sec.~\ref{sec:color}), \WHIZARD/ treats color in the
color-flow basis.  Therefore, a particle does not have a definite color state,
but it is part of zero or more color lines.  The color quantum numbers of a
particle are the color line indices in which it participates.  An amplitude is
always diagonal in color, if defined this way.  If we support only colorless
particles (which include $U(1)'$ ghost gluons), quarks, antiquarks, and
gluons, there are at most two color indices per particle.  Once an amplitude
is squared, color is either summed over, including interferences, or projected
onto definite color.  Therefore, squared amplitudes can use the same
representation as amplitudes.

For helicity, we have to select a specific basis.  The choice made for \OMEGA/
and \WHIZARD/ calculations is detailed in App.~\ref{sec:spinors}.  \OMEGA/ and
\WHIZARD/ deal with helicity amplitudes, therefore an amplitude is diagonal
(but correlated) in helicity.  In squared amplitudes -- spin density matrices
--, quantum entanglement must be supported, so each particle has two helicity
entries (bra and ket).

\subsubsection{Interactions}

An interaction object is an extension of the state-matrix object.  In addition
to the state matrix for $n$ particles, it contains a list of $n$ corresponding
momenta.  As stated above, the latter are well-defined, so the amplitude array
is still associated to the intrinsic quantum numbers.  

The \texttt{interaction} data type separates its particles into incoming,
virtual, and outgoing particles, and it establishes a parent-child relation
between them.  Furthermore, for each particle, it may contain a reference to a
corresponding ``source'' particle in another interaction, implemented as a
\FORTRAN/ pointer.

Hence, the \texttt{interaction} data type enables the program to represent a
physical event or process, broken down into proper subprocesses, in a
completely generic way, including full quantum correlations.  In practice, a
typical event consists of the \texttt{beam} interaction which has the colliding
particles outgoing, interactions representing structure-function applications
including radiation, the hard interaction, decays of the final state
particles, and possibly more.  These are represented as \texttt{interaction}
objects with appropriate pointers linking them together.

When events are ready for writing them to file, simulating actual
events in an experiment, entanglement and correlations must be resolved.
\WHIZARD/ provides methods for factorizing the correlated state into
one-particle states in different modes: averaging-out helicity, projecting
onto definite helicity for each particle, or keeping a one-particle spin
density matrix for each particle.  (The latter method is currently unsupported
by the standard event output formats, but available internally.)

\subsubsection{Evaluators}

When a physical event is constructed, the amplitude entries in the component
interactions must be squared and multiplied in a particular way.  For
instance, for the squared matrix element proper -- the sampling function for
the Monte Carlo integration -- the hard interaction must be squared,
convoluted with beam structure functions, and summed or averaged over
intrinsic quantum numbers.

This procedure is guided by the quantum number assignments and relations
between the various interactions.  Since the quantum numbers are static,
identical for all events of a specific type, but kinematics and the numeric
amplitude entries vary from event to event, it is advantageous to do the
bookkeeping only once.

The \texttt{evaluator} data type is an extension of the \texttt{interaction}
type.  In addition to the quantum numbers and momenta representing an
interaction, it holds a multiplication table together with suitable pointers
to one (squaring) or two (multiplication/convolution) source
interactions.   When an event is evaluated, the individual interactions are
first filled by momenta and amplitude values, then the corresponding
evaluators are activated by processing their multiplication tables.  The
result is a final evaluator that holds the complete event as an interaction
together with the entries of the final density matrix, suitably averaged or
summed over quantum numbers.

A \WHIZARD/ process implements three distinct objects as final evaluators: one
for the squared amplitude summed over everything, used for integration, and
two additional evaluators used for simulation: one differential in helicities
(as far as necessary), summed over color including interferences, suitable for
applying decays, and one differential in helicities and colors, suitable for
tracking color information.

\subsection{Event Generation}

With \WHIZARD/, simulated events can be generated after several
adaptive iterations have resulted in reasonably stable \VAMP/
integration grids, and a number of final iterations have yielded a
reliable estimate for the total cross section. The \VAMP/ grids are
then fixed, and an arbitrary number of further events is
generated. For each event, a first random number selects one of the
possible channels, taking its relative weight into account, and within
this channel a point is selected by importance sampling, taking the
adapted binning into account.  The event is kept or rejected according
to the ratio of the integrand at this point compared with the 
channel-specific maximum weight.  This results in a sequence of events
that simulate an actual experiment.  

Alternatively, for plotting distributions with greater accuracy, the
weighted events can be recorded as-is.

Since the estimate for the maximum weight can only be determined by
statistical sampling, the reweighting -- like any other statistical
method -- cannot exclude that the integrand 
value for a particular event exceeds this maximum estimate.  This
could be taken into account by again reweighting the whole sample
according to the new maximum estimate.  However, since \WHIZARD/ is
set up to put out unweighted events directly, we have chosen to merely
record these excess events and to compute, at the end, the value of
the error introduced by this excess.  It turns out that in practice,
this error is sufficiently below the overall integration error and can
be ignored -- if desired, it is possible to plot distributions of
excess events and check for critical regions where the adaptation
process could have failed.

\subsection{Decays}

\WHIZARD/~2 supports (cascade) particle decays in all simulated event
samples.  To enable this, both the production process and the required decay
processes have to be declared, compiled, and integrated over phase space, so
\VAMP/ grids are available for event generation.  Any massive particle species
can be declared as unstable, specifying its allowed decay channels.  During
simulation, \WHIZARD/ will scan over those particles and generate decay events
for them iteratively, until a set of stable particles is reached.

Technically, this involves cloning the process objects for the decay processes
and concatenating their evaluators event by event.  If more than one decay
channel is possible, the actual decay chain is selected on the basis of
random-number generator calls, distributed proportional to the respective
partial decay widths.

The use of \texttt{evaluator} objects for cascade decays ensures that all
color and spin correlations are kept.  The program always computes
color-summed and color-projected matrix elements separately.  The color-summed
matrix element, which is exact (at tree level), determines the decay angle
distribution of the final state particles.  Internal helicities are summed
over only after convoluting the matrix elements, and final helicities can be
kept if desired.  The color-projected matrix element is then used to determine
the color flow in the $1/N_c$ approximation, based on the relative
probabilities of all flows allowed for the particular decay chain with the
selected kinematics.

In the integration step, and in the simulation of stable-particle events,
initial and final state are completely specified (up to a possible summation
over equivalent massless particles), so this mode generates exclusive final
states.  When decays are enabled, all final states accessible by the decay
chain can be produced; this corresponds to a more inclusive treatment of
particle production.  If the final state is identical, the comparison of the
exclusive calculation with complete matrix elements and the factorized
decay-chain calculation reveals the effects of off-shell intermediate states
and irreducible background diagrams.  Note that complete matrix elements and
on-shell factorization both respect gauge invariance, while restricting an
exclusive matrix element to specific (off-shell) intermediate states or
Feynman graphs, also supported by \WHIZARD/, may lead to gauge-dependent
results.

For illustrating the effects of spin correlations, the \texttt{unstable}
declaration allows, for each unstable particle separately, to request either
full spin correlations in its decay, classical correlations only (diagonal
density matrix), or no correlations, i.e., isotropic decay.

\subsection{Interfaces}

So far, we have described \WHIZARD/ as an event generator that is
able, for fixed collider energy, to compute the partonic cross section
for a scattering process, or a partial width for a decay process, and
to generate simulated partonic events for this process.  Actually,
while the adaptation and integration proceeds separately for each
process selected by the user, in the event generation step an
arbitrary number of processes can be mixed.

For a complete physics simulation, this is not sufficient.  First of
all, in realistic colliders the partonic c.m.\ energy is not fixed.
At hadron colliders, this energy is distributed according to parton
distribution functions (PDFs).  At lepton colliders, it is distributed
according to the beam energy spectrum, affected mostly by
beamstrahlung.  Furthermore, initial-state radiation (ISR) reduces the
available partonic energy.  To account for this, \WHIZARD/ is able to
include the partonic energy spectrum in the integration.  Each
spectrum or radiation effect introduces an extra energy variable and
thus increases the integration dimension by one.  Since several
effects may have to be convoluted (e.g., beamstrahlung + ISR), the
number of extra integrations may be larger than two.

For computing these effects, \WHIZARD/ makes use of external programs
and libraries.  While electromagnetic ISR is accounted for internally,
for beamstrahlung and photon-collider spectra there are two options:
the \CIRCE/1/\CIRCE/2~\cite{circe} packages are now also contained in
the \WHIZARD/ bundle.  To account for generic $e^+e^-$ energy spectra,
\WHIZARD/ can read events from \GUINEAPIG/~\cite{guineapig} output.
PDFs are taken from the standard \LHAPDF/~\cite{lhapdf} library.

Parton-shower, i.e., QCD radiation is not yet accounted for internally
by \WHIZARD/.  However, \WHIZARD/ respects the Les Houches
Accord~\cite{Boos:2001cv} and therefore can interface to parton-shower
Monte Carlo programs.  To this end, events should be written to file in \LHEF/
format.  The events can then be treated by shower generators such as \PYTHIA/.
This allows not just for showering partons
(assuming that double-counting is excluded, i.e., the hard \WHIZARD/
process does not include parton radiation), but also for interfacing
hadronization, underlying events, etc. On the other hand, the
infrastructure for an own parton shower generator is already
included.

\section{User Interface}

\subsection{Installation and Prerequisites}
\label{sec:installation}

The \WHIZARD/ package is available as a \texttt{.tar.gz} file\footnote{The
  package is designed for UNIX systems, LINUX and MacOS in particular.  Other
  operating systems may also be supported in the future.} via the
  HepForge page:
\begin{center}
  \url{http://www.hepforge.org/downloads/whizard}
\end{center}
This includes several auxiliary packages (\OMEGA/ matrix element generator,
\VAMP/ integration, \CIRCE/ beamstrahlung, etc.).  Two compilers are needed:
(i) a \FORTRANOBJECTS/ compiler, alternatively a \FORTRANNINETYFIVE/ 
compiler with support for selected \FORTRANOBJECTS/ features\footnote{Consult
  the \WHIZARD/ website or contact the authors for the current support status
  of \FORTRAN/ compilers.} for \WHIZARD/; (ii) the \OCAML/
  compiler\footnote{\OCAML/ is part of most standard LINUX distributions;
  otherwise it is available free of charge from
  \url{http://caml.inria.fr/ocaml/}.}~\cite{O'Caml} for \OMEGA/.

For hadron-collider applications, the \LHAPDF/ parton-distribution function
library should be available on the system when \WHIZARD/ is configured.  The
same holds for \HEPMC/ or \STDHEP/ (event-file formats), if these features are
needed.

\WHIZARD/ and its subpackages are set up following \texttt{autotools}
conventions.  The package is configured by \texttt{configure} and built and
installed by \MAKE/ commands.  The default installation path is
\texttt{/usr/local}, but different installation locations can be selected by
the usual \texttt{configure} options.  The installation process results in a
single executable \texttt{whizard} which is located, by default, in
\texttt{/usr/local/bin}.  Auxiliary files will be installed in
\texttt{/usr/local/lib/whizard} and \texttt{/usr/local/share/whizard}.
Alternatively, non-default installation paths can be selected by
standard \texttt{configure} options such as \verb|--prefix|.

\subsection{\SINDARIN/}

The \WHIZARD/ executable program takes its input from a script, which can
either be executed interactively, or read from a file.\footnote{A \C/-compatible
API that allows for treating \WHIZARD/ as an external library is
planned for a future revision.}  The script is written
in a domain-specific language called \SINDARIN/\footnote{Scripting
  INtegration, Data Analysis, Results display, and INterfaces}.  All input
needed for the Monte Carlo run -- choice of model, processes, beams,
parameters, cuts, etc. -- is specified within a \SINDARIN/ script.

\SINDARIN/ is a complete programming language, designed to suit the needs of
Monte-Carlo integration and simulation.  On the top level, a \SINDARIN/
script consists of commands that steer the execution of the Monte Carlo.
Examples are: \texttt{integrate}, \texttt{simulate}.  The commands take
arguments, for instance
\begin{quote}
  \texttt{integrate (prc\_tt)}
\end{quote}
where \texttt{prc\_tt} is an identifier for the (partonic) process to
integrate, and possibly optional arguments:
\begin{quote}
  \texttt{integrate (prc\_tt) }\verb|{| \texttt{iterations = 5:10000} \verb|}|
\end{quote}
Some commands take the form of assignments, in particular the command that
defines a process and declares its identifier
\begin{quote}
  \texttt{process prc\_tt = g, g =}\verb|>| \texttt{t, tbar}
\end{quote}
or a beam declaration command
\begin{quote}
  \texttt{beams = p, p =}\verb|>| \texttt{lhapdf}
\end{quote}
which, in the example, also declares that \LHAPDF/ parton distribution
functions are to be used.

\SINDARIN/ supports variables, both predefined variables (such as particle
masses) and user-defined variables.  Variables are typed.  The available types
are logical, integer, real, complex, string, particle alias (e.g., \texttt{q =
  u:d}) and subevent.  Variables, constants, operators and functions operating
on them build expressions.  There are the usual arithmetic and string
expressions.  Furthermore, \SINDARIN/ supports expressions that involve
particle aliases and subevents.  They can describe observables, trigger and
cut conditions of a rather generic kind, to be applied to integration and
simulation.

The language contains constructs that enable data visualization.  Commands and
expressions can be evaluated based on conditions, and further script files can
be included.  There is also a loop construct that allows for scanning over
parameters.

The \SINDARIN/ language, and thus the \WHIZARD/ user interface, is described
in detail in the \WHIZARD/ manual,
\url{http://projects.hepforge.org/whizard/manual.pdf}.

\subsection{Implementation of the Language}

The conception of a programming language as a replacement for fixed-format
input files requires the implementation of lexer, parser, and compiler (or
interpreter) for this language.

These implementations are done in a generic way, so arbitrary syntax
structures can be handled, and the \SINDARIN/ syntax is just a special case.
Actually, \WHIZARD/ processes a few additional, albeit much simpler, syntax
structures (e.g., the model-file syntax, the SLHA syntax) using the same lexer
and parser implementation.

The lexer analyzes an input stream, which may come from an external or
internal file or string, and separates it into tokens.  It has a basic notion
of data types, so it distinguishes numerical values from string identifiers,
and it identifies keywords from a given syntax table.  Furthermore, it can
handle comments, matching delimiters and matching quotes.  The lexer
definition assigns characters to appropriate character classes.

Syntax tables are coded in form of a table of strings that are not necessary
hard-wired.  The table entries are equivalent to a formal syntax description
which declares each syntax element as atomic, alternative, or sequence, with
some specific variants that describe frequent cases.  Sequence and alternative
elements are defined in terms of other syntax elements.  Each syntax table is
checked at runtime for completeness and consistency.

The parser is implemented as a simple top-down parser.  The input, as a
sequence of tokens, is matched element by element against the syntax table.
If a syntax element matches, the token is inserted into a growing parse tree.
If not, it is put back into the token stream.

The \SINDARIN/ compiler is split into two parts.  Expressions (of any type)
are compiled into an evaluation tree.  In this step, constant expressions are
evaluated immediately.  To have a compiled version is useful for cut
expressions in particular, since they are evaluated once for each event during
integration and simulation.  The other part of the compiler handles commands
and assignments.  The corresponding parse-tree elements are transformed into
objects that collect the relevant data; command execution then amounts to
calling an ``execute'' method on the object.


\subsection{Physics Models}
\label{sec:models}

The physics model to be used for process definitions is declared in the
\SINDARIN/ script, for instance: \texttt{model = MSSM}.  It is possible to use
several models concurrently for distinct processes in a single script.

The support for specific models in \WHIZARD/ relies on the
implementation of the corresponding models in \OMEGA/ and \WHIZARD/. In
both packages, the infrastructure supports the incorporation of
particles with spin 0, $\frac12$ (Dirac/Majorana fermions), 1,
$\frac32$ and 2. Since the structure of \OMEGA/ allows for the
incorporation of arbitrary higher-dimensional operators all possible
physics models based on quantum field theories containing particles
with spins up to two can be implemented; even more general setups are
possible like models based on noncommutative generalizations of
space-time (see below). 

Specific physics models are defined with the help of their particle
content (and the corresponding quantum numbers), the fundamental
interactions (vertices) and -- most importantly and error-prone -- the
set of coupling constants and parameters together with the relations
among them. Within \OMEGA/, some basic toy models
like QED and QCD as well as the SM and its
derivatives (like non-unitary gauges, extensions with anomalous
couplings with and without $K$ matrix unitarization, non-trivial CKM
matrix) are implemented in \texttt{models.ml}, the supersymmetric
models like the MSSM and possible extensions (NMSSM, PSSSM etc.) in
\texttt{modellib\_MSSM.ml}, \texttt{modellib\_NMSSM.ml}, and
\texttt{modellib\_PSSSM.ml}, while non-SUSY BSM extensions (like Little
Higgs models, $Z'$ models and extra dimensional models) are
implemented in \texttt{modellib\_BSM.ml}. In this module there is also
a model \texttt{Template} which has exactly the same content as the
SM, but can be augmented by the user to incorporate new particles and 
interactions in his or her favorite model. More details about how
this works can be found in subsection~\ref{sec:buildmodels}.

\begin{table}
  \begin{center}
  \begin{tabular}{|l|l|l|}
    \hline
    Model type & with CKM matrix & trivial CKM \\
    \hline
    QED with $e,\mu,\tau,\gamma$ & --- &  \tt{QED} \\
    QCD with $d,u,s,c,b,t,g$ & --- &  \tt{QCD} \\
    Standard Model        & \tt{SM\_CKM} & \tt{SM} \\
    SM with anomalous couplings &  \tt{SM\_ac\_CKM} &
    \tt{SM\_ac} \\
    SM with charge -4/3 top & --- & \tt{SM\_top} \\
    SM with anomalous top coupl. & --- & \tt{SM\_top\_anom} \\
    SM with $K$ matrix &  \tt{SM\_km\_CKM} &
    \tt{SM\_km} \\
    SM with triangle Higgs coupl. &  --- &
    \tt{SM\_triangle\_higgs} \\
    \hline
    SUSY Yang-Mills & --- & \tt{SYM} \\
    MSSM &   \tt{MSSM\_CKM} & \tt{MSSM} \\
    MSSM with gravitinos & ---  & \tt{MSSM\_Grav} \\
    NMSSM &   \tt{NMSSM\_CKM} & \tt{NMSSM} \\
    PSSSM &   --- & \tt{PSSSM} \\
    \hline
    Littlest Higgs &  --- & \tt{Littlest} \\
    Littlest Higgs with ungauged $U(1)$ &  --- &
    \tt{Littlest\_Eta} \\
    Littlest Higgs with $T$ parity &  --- &
    \tt{Littlest\_Tpar} \\
    Simplest Little Higgs (anomaly-free) &  --- &
    \tt{Simplest} \\
    Simplest Little Higgs (universal) &  --- &
    \tt{Simplest\_univ} \\
    SM with spin-2 graviton & --- & \tt{Xdim} \\
    SM with gravitino and photino & --- & \tt{GravTest} \\
    \hline
    SM with generic $Z'$ & --- & \tt{Zprime} \\
    Universal Extra Dimensions & --- & \tt{UED} \\
    3-site model & --- & \tt{Threeshl} \\ 
    3-site model without heavy fermions & --- & \tt{Threeshl\_nohf} \\ 
    \hline
    Augmentable SM template & --- & \tt{Template} \\
    \hline
  \end{tabular}
  \end{center}
  \caption{List of models that are currently supported by \WHIZARD/:
    the SM and its relatives, simple subsets of the SM,
    the MSSM, other models beyond the SM as well as a template which
    can be augmented by the user to include additional new particles
    and interactions.} \label{tab:models}
\end{table}

In \WHIZARD/ for each model \texttt{MODEL} there is a file
\texttt{MODEL.mdl} which contains the particles with their quantum
numbers (electric charge, color etc.) as well as a definition of the
basic parameters that can be accessed via the input file. This file
also contains a list of all the vertices of the model, which is
important for the generation of phase space of processes in that
specific model. For each model \texttt{MODEL}, there is also a file
\texttt{parameters.MODEL.f90} which contains all the couplings
of the corresponding model as functions of the basic input
parameters. An overview over the publicly supported models as well as
those currently in their testing phase are shown in
Table~\ref{tab:models}.


\subsection{Processes}

For a given physics model, \WHIZARD/ can compute cross sections or partial
decay widths for all processes that are physically allowed.  The
user-specified list of processes can be arbitrary, as long as the computer is
capable of dealing with it.\footnote{Typical bottlenecks are: complexity of
  the matrix element (CPU time), complexity of phase space (memory), number of
  contributing subprocesses (both).}  For each process, the \OMEGA/
matrix-element generator generates a tree-level matrix element, so without
manual intervention, the result corresponds to fixed leading order in
perturbation theory.

To define a process, the user may completely specify incoming and outgoing
particles, choosing from the elementary particles contained in the selected
model.  For convenience, it is possible to define particle aliases and to sum
over massless particles in the incoming or outgoing state, e.g., combine all
neutrino generations or all light quarks.  In this case, all contributing
matrix elements will be added at each selected phase-space point, and the code
generated by \OMEGA/ is able to take advantage of cross-flavor common
subexpression elimination.  For the generated events, a particle combination
will be selected event by event according to the relative weight of the
corresponding squared matrix element.

The user can restrict intermediate states to select or exclude classes of
Feynman graphs.

\subsection{Beams and Partons}

Once the processes have been declared, all processes in the
are available for calculating cross sections (with
arbitrary cuts) and simulating events.

The user selects a list of processes among the available ones and
specifies the type and energy of the colliding beams (or the type of
decaying particle).  Each beam can be given a structure or
polarization.  For instance, in hadron collisions, the beam structure
is given by the PDF set, specified by the usual \LHAPDF/
parameters. Lepton collisions are affected by beamstrahlung and
electromagnetic initial-state radiation.  Photon collisions proceed
via \CIRCE/2 spectra or via photons radiated from leptons in the
effective-photon approximation.  In all cases, all free parameters can
be set and modified in the input file.

Apart from these physical beam setups, it is possible to compute
fixed-energy cross sections, partial widths, and event samples for
any types of colliding or decaying particles.

In lepton and photon collisions, polarization is of importance.  For
each beam, the user can specify the longitudinal polarization or,
alternatively, transversal polarization.  Furthermore, it is possible
to specify a complete spin-density matrix for the incoming beams. 
Since helicity amplitudes are used throughout the program, the
polarization of final-state particles can also be extracted.

\subsection{Parameters, Cuts, and Other Input}

\WHIZARD/ follows the philosophy that no numerical parameters are
hard-coded, everything can be specified by the user in the input script.
However, wherever applicable, reasonable default values exist.  With the caveat
that some parameter relations are fixed by the model definition (to
ensure gauge invariance), all free physics parameters such as particle
masses, widths, and couplings can be modified in the input file.  This
also implies that the phase-space setup, which depends on particle
masses, is generated afresh for each \WHIZARD/ run.

For supersymmetric models, there is the SLHA standard~\cite{slha} which
specifies how to transfer physics parameters between programs.
There is a specific \SINDARIN/ command that reads in a SLHA file.

Cuts on phase space are of particular importance.  Many cross sections are
infinite if no cuts are applied.  To avoid confusion, \WHIZARD/ by default
does not apply any cuts, so ensuring a finite cross section is entirely left
to the user.  However, it is rather simple to define generic cuts that render
all integrations finite (e.g., cutting on $p_T$, rapidity, and separation of
all visible particles).  For specifying user cuts, a wide range of observables
such as energy, $p_L$, $p_T$, angles, rapidity, etc.\ is available.  Cuts are
defined by applying observables or expressions involving observables to events
or subevents selected by user-defined criteria.

All parameters (in fact, all commands) can also be set on the command line.
This facilitates the use of \WHIZARD/ in shell scripts.

\subsection{Using and Analyzing Results}

The \WHIZARD/ user interface has been designed with various
applications in mind, ranging from theoretical studies to experimental
simulation.

A theoretical study typically implies the calculation of some cross
section and the display of characteristic distributions of
observables.  To this end, the user would set up the
processes and parameters, run the program to compute cross section
integrals, and generate a sufficiently large sample of weighted
events.  In this case, one would not use the rejection algorithm to
unweight events, so no information is lost.  It is possible to write
the event sample to file and to do analyses by some external program,
but \WHIZARD/ also contains its own analysis module.  With this
module, the user specifies lists of observables to histogram (on top
of, and analogous to specifying cuts). During event generation,
the program will fill those histograms and output data tables.  To
plot such data, \WHIZARD/ employs the \GAMELAN/ package.  This program
generates encapsulated \POSTSCRIPT/ code that can conveniently be
included in \LaTeX\ documents.

For a simulation study, the user needs a sequence of unweighted,
fully hadronized events.  The \WHIZARD/ run includes the necessary
steps of adaptation and integration and proceeds to the generation of
unweighted events; the event sample may be specified either by the
number of events or by an integrated luminosity.  Hadronization is
accomplished by linking \PYTHIA/ or some other hadronization package
to \WHIZARD/, preferably by reading an event file that \WHIZARD/ has written
in the \LHEF/ standard.  \WHIZARD/ supports several event file formats,
including the \STDHEP/ binary format.  These event samples are ready to be
further processed by detector simulation and analysis.

It is often necessary to re-run a program several times in order to
change or refine event numbers, analysis parameters, etc.  Since
adaptation, integration, and event generation all can take
considerable time, \WHIZARD/ provides means for reusing unfinished or
previous integration results, grids, and events, so the program needs
not start from the beginning.  The integrity of data is checked by MD5
sums.  Furthermore, \WHIZARD/ is able to rescan or reprocess event files
produced by other programs, if they are available in \HEPMC/ format.  This is
useful for computing, e.g., exact matrix elements for reweighting Monte-Carlo
samples. 

\section{Extensions and Extensibility}


\subsection{Building Models}
\label{sec:buildmodels}

If a model can be formulated for the
\FEYNRULES/~\cite{Christensen:2008py} package, it can be made
automatically available to \WHIZARD/. A specific
interface is available for both versions, \WHIZARD/1 and
\WHIZARD/2. For \WHIZARD/2 the interface is very convenient, as there
is a plugin mechanism which directly incorporates the models into the
main program, such that models generated via \FEYNRULES/ can be used
in the same way as those hard-coded in the program core. For more
details about the interface as well as physics examples confer the
specific publication~\cite{Christensen:2010wz}.

If the \FEYNRULES/ capabilities are not sufficient, adding a new model
to \WHIZARD/ is nevertheless straightforward. To manually add a new
model, one has to edit both \WHIZARD/'s model file and the \OMEGA/
driver simultaneously. In the file \texttt{modellib\_Template.ml} in
the \texttt{src} directory of \OMEGA/ there is a \texttt{Template}
module which is just a copy of the SM implementation within
\OMEGA/. From this template one can read off the syntax structure and
add new particles and interactions according to one's favorite new
physics model. This exhausts the changes that have to be made on the
\OMEGA/ side.

The next step is to add all new particles with their quantum numbers
in the file \texttt{Template.mdl} in the subdirectory
\texttt{share/models} of \WHIZARD/. In the bottom part of 
that file all new interaction vertices have to be added in the
way of the SM vertices already written down there. This is important
in order that \WHIZARD/ can find the phase space channels including
the new particles added by the user. The hardest and most error-prone
work to do is to add the functional relations among the coupling
constants and parameters beyond the SM within the corresponding
parameter file \texttt{parameters.TEMPLATE.f90} in the
directory, \texttt{src/models}. Again, the examples from
the SM might serve as a guideline here for the way how to incorporate
all the new couplings in this file. The model \texttt{Template} can be
accessed in \WHIZARD/ with the tag \texttt{Template} in the same way
as the other models defined in subsection \ref{sec:models}. It can
even be used when the user has not added any new particles or
interactions; in that case it is just a mirror of the SM.

\subsection{Improving or Replacing Matrix Elements}
\label{sec:hacking}

The matrix-element source code generated by \OMEGA/ is very easy to read and
consequently also to modify.   In Appendix~\ref{sec:sample-code}, we show
the complete $e^+e^-\to\mu^+\mu^-$ scattering amplitude in the SM.
Notice that, for convenience, the crossed amplitude with all particles
outgoing is calculated internally.  For this reason the incoming
momenta are reversed.

In the code, \verb|mass| is an array of particle masses, indexed by
the PDF Monte Carlo particle codes, that is defined in the module
\verb|omega_parameters|.  \verb|qlep|, \verb|gnclep(1)| and
\verb|gnclep(2)| are the lepton charge and vector and axial vector
neutral current coupling constants, respectively.  \verb|wd_tl| is a
function that returns a non-zero width for time-like momenta.  The
functions \verb|pr_feynman| and \verb|pr_unitarity| implement the
propagators and the functions \verb|v_ff| and \verb|va_ff| implement
vector couplings and mixed vector/axial couplings of the fermions
given as arguments.

It is now straightforward to replace any of these functions by another
function that computes a non-standard propagator or coupling or to add
another particle exchange, like a~$Z'$.  Of course, it is more
efficient for a comprehensive analysis of a $Z'$-model to produce a new
model file, but non-standard vertices are a useful hook for adding
radiative corrections (see sec.~\ref{sec:nlo}).  When preparing
modified vertex factors for fermions, it is most convenient to use the
elementary vertex factors for the $\gamma$-matrix structures, as they
are already optimized and guaranteed to be consistent with the
conventions used in the other functions from \OMEGALIB/.

Note that the final line, which probably takes on a form like
\begin{displaymath}
  \verb|oks_l1bl1l2bl2 = - oks_l1bl1l2bl2|
\end{displaymath}
takes care of all the factors of~$\ii$ coming from vertices and
propagators in the Feynman rules.  Any modification of the amplitude
must respect this convention, in order not to spoil potential
interference terms.

The normal workflow would let \WHIZARD/ recompile and relink matrix-element
source code only if the process declaration had changed.  With a
\verb|--recompile| flag set (or the \SINDARIN/ parameter
\verb|?recompile_library|), the
modified file will be treated by the program as if it was the originally
generated code.  Clearly, to prevent accidental overwriting a modified file,
it should be additionally saved in a place different from the current working
directory.

\subsection{Higher Orders}
\label{sec:nlo} 

To match the experimental precision of hadron and lepton collider
environments, theoretical predictions have to include higher order
radiative corrections, originating from virtual and real
diagrams. 

The precise meaning of ``higher order'' or ``next-to-leading order'' (NLO)
very much depends on the context.  In lepton-collider physics, the most
important part is usually QED radiation, since this effect results in
infrared and collinear divergences.  They can partly be analytically treated
and resummed.  \WHIZARD/ accounts for higher-order radiation via the
well-known ISR structure function which can be activated when appropriate.

A refinement of the NLO treatment typically involves a complete one-loop
calculation in the SM, or one of its (perturbatively tractable) extensions.
For instance, in~\cite{omwhiz_nlo} precompiled NLO matrix elements
for the production of two SUSY particles (charginos) 
at the ILC have been used and linked to \WHIZARD/~1 in the form
of an external matrix element, convoluted with a
user-defined structure function.  Since this calculation was
performed via the automatic \texttt{FeynArts} -- \texttt{FormCalc} --
\texttt{LoopTools} toolchain~\cite{fa_fc_lt}, the result showed that
\WHIZARD/ can be extended to a NLO event generator for lepton-collider
processes, including complete electroweak and supersymmetric
corrections. However, there
is no automatic implementation yet.

For hadron colliders such as the LHC, the numerically
  dominant higher-order corrections typically are pure QCD
  corrections, originating from radiation and loops involving
gluons and massless quarks.  Multiple QCD radiation from
both initial and final-state partons involves disparate energy scales and has
to be matched to both the 
hard process and to non-perturbative models of hadronization and
multiple interactions.
The current \WHIZARD/ version does
not yet address QCD beyond tree level.  For the parton shower, it
provides an independent algorithm and implementation that will be
described in a separate publication~\cite{whizard-shower}; 
alternatively, the user
can use the well-established \PYTHIA/~\cite{pythia} code for generating QCD
radiation, either externally or automatically from 
within the program.  Leading-order parton-shower matching is available
in form of the MLM algorithm~\cite{MLM}.  Implementations of dipole
subtraction, interleaved parton-shower and parton-shower
matching, and further QCD 
effects are under development and will be merged into the \WHIZARD/
framework.  Thus, the necessary ingredients for consistently
simulating QCD beyond leading order are projected as intrinsic parts
of the program in a future release.

\section{Conclusions and Outlook}

Data taking at the LHC has begun and almost the whole standard model
has already been rediscovered at the time of writing. At the same
time, the physics and detector studies for the planned ILC are being
refined with increasing requirements on the accuracy of theoretical
predictions.  In both cases Monte Carlo simulation tools must respond
to the challenge to provide a flexibility and theoretical accuracy
that will enable us to uncover the true nature of physics in the
$\mathrm{TeV}$ energy range.

Event generators with complete multi-particle matrix elements
at the hard-interaction level are not designed to completely
replace well-established tools
that simulate few-particle production and subsequent
decays.  Nevertheless, they already have proven indispensable for
refining the accuracy of predictions, simulating complex elementary
processes, and providing reliable background estimates where data
alone are insufficient for unambiguous signal extraction.

Version~1 of \WHIZARD/ had been designed for ILC studies with no
colors in the initial state and a moderate number of colored jets.  As
a result, color was not built in from the beginning and the
implementation was not optimal.  This has changed dramatically with
the redesigned Version~2 of \WHIZARD/, where QCD in the color flow
representation has been built in from the ground up. This new version
propels \WHIZARD/ into the LHC era.

Simultaneously, the streamlined architecture of Version~2 of \WHIZARD/
provides for a much simpler installation and usage of the program.
The executable can be installed in a central location and is
controlled by a single input file describing the analysis in a
flexible language, close to physics.  This also allows for easy
snapshots of the installation for later verifications.

To show the applicability of \WHIZARD/ for high-multiplicity hard
interactions in LHC processes we calculated cross sections for 
multi-parton processes of the Drell-Yan type $pp \to (W \to \ell\nu) + n \,
j$ with the number of jets equal to $n=1,2,3,4,5$, as
well as multi-parton process associated to top/Higgs production
and background $pp\to \ell\ell\nu\nu b\bar b+n\,j$ with $n=0,1,2$.  
These calculations have been performed in the complete
Standard Model: jets include gluons
together with four light quark flavors, and the matrix
elements incorporate all interactions that involve photon and weak boson
exchange, adding to and interfering with the QCD part.
We observe that \WHIZARD/ is able to simulate,
e.g., $W$+4 jet processes in a straightforward way;
computing $W$+5 jets is also possible with slightly more effort. 
Computing time and memory usage rise
roughly by about a factor of 10 for each additional jet that is added.
All processes considered in this paper are tractable
with standard current workstations, given up to a few GB
of memory, up to several days of adaptation/integration time, and up to a few
more days of CPU time for subsequently generating an unweighted event sample
that corresponds to $1\;\textrm{fb}^{-1}$ of LHC luminosity.

The new version of \WHIZARD/ provides a stable framework for further
developments: an improved matching of hard matrix elements with parton
showers, as well as other aspects of soft and collinear QCD, and the
fully automated incorporation of higher orders of perturbation theory.
The latter is particularly challenging and while \WHIZARD/ has already 
been used successfully in NLO
calculations~\cite{Binoth:2009rv,Greiner:2011mp}, many 
new techniques will have to be developed before the automated
construction of NLO event generators will have reached the same level
of maturity as in the LO case today.
In summary, \WHIZARD/ covers complex hard scattering processes in the
standard model and most of its known extensions at all past, current
and future high-energy colliders 
efficiently.  The program 
is ready for use as a universal and flexible tool for experimental
data analysis, data interpretation, and future phenomenological
studies.

\section*{Acknowledgements}

Special thanks go to the very recent \WHIZARD/ contributors,
F.~Bach, H.W.~Boschmann, F.~Braam, S.~Schmidt, D.~Wiesler, and especially
C.~Speckner. Furthermore, we would like to thank A.~Alboteanu,
T.~Barklow, M.~Beyer, T.~Binoth($\dagger$), E.~Boos, R.~Chierici,
K.~Desch, S.~Dittmaier, T.~Feldmann, T.~Fritzsche, N.~Greiner,
K.~Hagiwara, T.~Hahn, W.~Hollik, M.~Kobel, F.~Krauss, P.~Manakos,
T.~Mannel, M.~Mertens, N.~Meyer, K.~M\"onig, M.~Moretti, D.~Ondreka,
M.~Peskin, T.~Plehn, D.~Rainwater, H.~Reuter, T.~Robens,
M.~Ronan($\dagger$), S.~Rosati, A.~Rosca, J.~Schumacher,
M.~Schumacher, S.~Schumann, C.~Schwinn, T.~Stelzer, S.~Willenbrock,
and P.~Zerwas 
for valuable discussions, comments and help during this project. WK
and JR acknowledge the friendly atmosphere within and support by the
particle physics groups at the University of Karlsruhe and DESY,
Hamburg, and the Aspen Center for Physics, where a lot of this work
has been initiated. JR wants especially to thank the particle physics
group at Carleton University, Ottawa, where part of this work has been
completed, for their warm hospitality and lots of interesting
discussions.  WK expresses his particular gratitude for the warm
hospitality and support of the particle physics group at the
University of Urbana/Champaign.  We would like to extend particular
gratitude to C.~Schwinn for his work on $R_\xi$-gauge functors and
tests of gauge parameter independence in \OMEGA/ amplitudes, and also
for many helpful and enlightening discussions in an early stage of
\OMEGA/.

This work has been supported in part by the Helmholtz-Gemeinschaft
under Grant No.{} VH-NG-005, the Helmholtz alliance ``Physics at the
TeraScale'', the Bundesministerium f\"ur Bildung und
For\-schung, Germany, (05\,HT9RDA, 05 HA6VFB, 05\,H4WWA/2, 05\,H09PSE), the
Ministerium f\"ur Wissenschaft und Kultur of the state
Baden-W\"urttemberg, and the Deutsche For\-schungs\-ge\-mein\-schaft
(single projects MA\,676/6-1 and RE 2850/1-1 as well as by the
Graduier\-ten\-kolleg GK 1102 ``Physics at Hadron Colliders''). 

\appendix
\section{Conventions}
\label{sec:spinors}

In this appendix, 
 we collect some of the conventions used in \WHIZARD/ for the
calculation of helicity amplitudes and polarized off-shell wave
functions. The
matrix elements generated by \OMEGA/ call functions from \OMEGALIB/
that implement the following conventions.  It is therefore
straighforward to replace these conventions by another set, should the
need ever arise in specialized applications.


\subsection{On-shell wavefunctions}

\subsubsection{Dirac and Majorana fermions}


We use the two-component Weyl spinors
\begin{subequations}
\begin{align}
  \chi_+(\vec p) &=
     \frac{1}{\sqrt{2|\vec p|(|\vec p|+p_3)}}
     \begin{pmatrix} |\vec p|+p_3 \\ p_1 + \ii p_2 \end{pmatrix} 
     \qquad
  \chi_-(\vec p) &=
     \frac{1}{\sqrt{2|\vec p|(|\vec p|+p_3)}}
     \begin{pmatrix} - p_1 + \ii p_2 \\ |\vec p|+p_3 \end{pmatrix}
\end{align}
\end{subequations}
to construct the four-component Dirac or Majorana spinors:
\begin{equation}
  u_\pm(p) =
     \begin{pmatrix}
       \sqrt{p_0\mp|\vec p|} \cdot \chi_\pm(\vec p) \\
       \sqrt{p_0\pm|\vec p|} \cdot \chi_\pm(\vec p)
     \end{pmatrix}
     \qquad
  v_\pm(p) =
     \begin{pmatrix}
       \mp\sqrt{p_0\pm|\vec p|} \cdot \chi_\mp(\vec p) \\
       \pm\sqrt{p_0\mp|\vec p|} \cdot \chi_\mp(\vec p)
     \end{pmatrix}
\end{equation}

For the implementation of purely Dirac fermions, there are also
expressions for the conjugated spinors, which are not used in the
mixed Dirac/Majorana implementation. There the conjugated spinors are
constructed with the help of the charge-conjugation matrix.


\subsubsection{Polarization vectors}
\label{sec:polvec}

We use the following conventions for spin-1 particles:
\begin{subequations}
\begin{align}
  \epsilon^\mu_1(k) &=
    \frac{1}{|\vec k|\sqrt{k_x^2+k_y^2}}
      \left(0; k_z k_x, k_y k_z, - k_x^2 - k_y^2\right) \\
  \epsilon^\mu_2(k) &=
    \frac{1}{\sqrt{k_x^2+k_y^2}}
      \left(0; -k_y, k_x, 0\right) \\
  \epsilon^\mu_3(k) &=
    \frac{k_0}{m|\vec k|} \left({\vec k}^2/k_0; k_x, k_y, k_z\right)
\end{align}
\end{subequations}
and
\begin{subequations}
\begin{align}
  \epsilon^\mu_\pm(k) &=
     \frac{1}{\sqrt{2}} (\epsilon^\mu_1(k) \pm \ii\epsilon^\mu_2(k) ) \\
  \epsilon^\mu_0(k) &= \epsilon^\mu_3(k)
\end{align}
\end{subequations}
i.\,e.
\begin{subequations}
\begin{align}
  \epsilon^\mu_+(k) &=
     \frac{1}{\sqrt{2}\sqrt{k_x^2+k_y^2}}
        \left(0; \frac{k_zk_x}{|\vec k|} - \ii k_y,
                 \frac{k_yk_z}{|\vec k|} + \ii k_x,
                 - \frac{k_x^2+k_y^2}{|\vec k|}\right) \\
  \epsilon^\mu_-(k) &=
     \frac{1}{\sqrt{2}\sqrt{k_x^2+k_y^2}}
        \left(0;  \frac{k_zk_x}{|\vec k|} + \ii k_y,
                  \frac{k_yk_z}{|\vec k|} - \ii k_x,
                 -\frac{k_x^2+k_y^2}{|\vec k|}\right) \\
  \epsilon^\mu_0(k) &=
     \frac{k_0}{m|\vec k|} \left({\vec k}^2/k_0; k_x, k_y, k_z\right)
\end{align}
\end{subequations}

These conventions are similar to those used in \HELAS/~\cite{HELAS}, which are
\begin{subequations}
\begin{align}
  \epsilon^\mu_\pm(k) &=
     \frac{1}{\sqrt{2}} (\mp \epsilon^\mu_1(k) - \ii\epsilon^\mu_2(k) ) \\
  \epsilon^\mu_0(k) &= \epsilon^\mu_3(k)
\end{align}
\end{subequations}
with the same definitions as above. 

Note that these conventions do not fit the definitions of the spinor
wavefunctions defined in the last paragraph. In fact, they correspond
to a different quantization axis for angular momentum. So, when
constructing spin-3/2 wavefunctions out of those for spin 1/2 and spin
1, a different convention for the polarization vectors is used. 


\subsubsection{Polarization vectorspinors}
The wavefunctions for (massive) gravitinos are constructed out of
the wavefunctions of (massive) vectorbosons and (massive) Majorana
fermions:
\begin{subequations}
\begin{align}
\psi^\mu_{(u; 3/2)} (k) &= \; \epsilon^\mu_+ (k) \cdot u (k, +) \\
\psi^\mu_{(u; 1/2)} (k) &= \; \sqrt{\dfrac{1}{3}} \, \epsilon^\mu_+ (k)
        \cdot u (k, -) + \sqrt{\dfrac{2}{3}} \, \epsilon^\mu_0 (k) \cdot
        u (k, +) \\
\psi^\mu_{(u; -1/2)} (k) &= \; \sqrt{\dfrac{2}{3}} \, \epsilon^\mu_0 (k)
        \cdot u (k, -) + \sqrt{\dfrac{1}{3}} \, \epsilon^\mu_- (k) \cdot
        u (k, +) \\
\psi^\mu_{(u; -3/2)} (k) &= \; \epsilon^\mu_- (k) \cdot u (k, -)
\end{align}
\end{subequations}
and in the same manner for $\psi^\mu_{(v; s)}$ with $u$ replaced by
$v$ and with the conjugated polarization vectors. These gravitino
wavefunctions obey the Dirac equation, they are transverse and they
fulfill the irreducibility condition
\begin{equation}
        \gamma_\mu \psi^\mu_{(u/v; s)} = 0 .
\end{equation}

As mentioned above, one needs to use the same quantization axis for
spin 1/2 and spin 1 in order to construct the correct spin-3/2
states. The polarization vectors
\begin{subequations}
\begin{align}
  \epsilon^\mu_+(k) &=
     \frac{- e^{+\ii\phi}}{\sqrt{2}}
        \left(0; \cos\theta\cos\phi - \ii\sin\phi,
                 \cos\theta\sin\phi + \ii\cos\phi,
                 -\sin\theta \right)  \\
  \epsilon^\mu_-(k) &=
     \frac{e^{-\ii\phi}}{\sqrt{2}}
        \left(0; \cos\theta\cos\phi + \ii \sin\phi,
                 \cos\theta\sin\phi - \ii \cos\phi,
                 - \sin\theta \right) \\
  \epsilon^\mu_0(k) &=
     \frac{1}{m} \left(|\vec k|; k^0\sin\theta\cos\phi,
                                 k^0\sin\theta\sin\phi,
                                 k^0\cos\theta\right)
\end{align}
\end{subequations}
are used exclusively for this purpose. 


\subsubsection{Polarization tensors}

Spin-2 polarization tensors are symmetric, transversal and traceless
\begin{subequations}
\begin{align}
  \epsilon^{\mu\nu}_{m}(k) &= \epsilon^{\nu\mu}_{m}(k) \\
  k_\mu \epsilon^{\mu\nu}_{m}(k) &= k_\nu \epsilon^{\mu\nu}_{m}(k) = 0 \\
  \epsilon^{\mu}_{m,\mu}(k) &= 0
\end{align}
\end{subequations}
with $m=-2,-1,0,1,2$.
\begin{subequations}
\begin{align}
  \epsilon^{\mu\nu}_{+2}(k) &= \epsilon^{\mu}_{+}(k)\epsilon^{\nu}_{+}(k) \\
  \epsilon^{\mu\nu}_{+1}(k) &= \frac{1}{\sqrt{2}}
     \left(   \epsilon^{\mu}_{+}(k)\epsilon^{\nu}_{0}(k)
            + \epsilon^{\mu}_{0}(k)\epsilon^{\nu}_{+}(k) \right) \\
  \epsilon^{\mu\nu}_{0}(k)  &= \frac{1}{\sqrt{6}}
     \left(   \epsilon^{\mu}_{+}(k)\epsilon^{\nu}_{-}(k)
            + \epsilon^{\mu}_{-}(k)\epsilon^{\nu}_{+}(k)
            - 2 \epsilon^{\mu}_{0}(k)\epsilon^{\nu}_{0}(k) \right) \\
  \epsilon^{\mu\nu}_{-1}(k) &=  \frac{1}{\sqrt{2}}
     \left(   \epsilon^{\mu}_{-}(k)\epsilon^{\nu}_{0}(k)
            + \epsilon^{\mu}_{0}(k)\epsilon^{\nu}_{-}(k) \right) \\
  \epsilon^{\mu\nu}_{-2}(k) &= \epsilon^{\mu}_{-}(k)\epsilon^{\nu}_{-}(k)
\end{align}
\end{subequations}
Here the polarization vectors from~\ref{sec:polvec} are used.



\subsection{Propagators}

Note that the sign of the momentum for fermionic lines is always
negative because all momenta are treated as \emph{outgoing} and the
particle charge flow is therefore opposite to the momentum.

\begin{itemize}
  \item Spin 0:
    \begin{equation}
       \frac{\ii}{p^2-m^2+\ii m\Gamma}\phi
    \end{equation}

  \item Spin 1/2:
    \begin{equation}
       \frac{i(-\fmslash{p}+m)}{p^2-m^2+\ii m\Gamma}\psi \qquad\qquad
       \bar\psi \frac{i(\fmslash{p}+m)}{p^2-m^2+\ii m\Gamma}
    \end{equation}
    The right one is only used for the pure Dirac implementation. 

  \item Spin 1 (massive, unitarity gauge):
    \begin{equation}
       \frac{\ii}{p^2-m^2+\ii m\Gamma}
          \left( -g_{\mu\nu} + \frac{p_\mu p_\nu}{m^2} \right) \epsilon^\nu(p)
    \end{equation}

  \item Spin 1 (massless, Feynman gauge):
    \begin{equation}
       \frac{-i}{p^2} \epsilon^\nu(p)
    \end{equation}

  \item Spin 1 (massive, $R_\xi$ gauge):
    \begin{equation}
       \frac{\ii}{p^2}
          \left( -g_{\mu\nu} + (1-\xi)\frac{p_\mu p_\nu}{p^2} \right)
       \epsilon^\nu(p)
    \end{equation}

  \item Spin 3/2:
    \begin{equation}
     \dfrac{\ii\biggl\{(-\fmslash{p} + m)\left(-\eta_{\mu\nu} + \dfrac{p_\mu
     p_\nu}{m^2}\right) + \dfrac{1}{3} \left(\gamma_\mu -\dfrac{p_\mu}{m}\right)
     (\fmslash{p} + m)\left(\gamma_\nu -
     \dfrac{p_\nu}{m}\right)\biggr\}}{p^2 - m^2  + \ii m
     \Gamma} \; \psi^\nu
    \end{equation}

  \item Spin 2:
    \begin{subequations}
    \begin{equation}
       \frac{\ii P^{\mu\nu,\rho\sigma}(p,m)}{p^2-m^2+\ii m\Gamma} T_{\rho\sigma}
    \end{equation}
    with
    \begin{multline}
       P^{\mu\nu,\rho\sigma}(p,m)
        = \frac{1}{2} \left(g^{\mu\rho}-\frac{p^{\mu}p^{\rho}}{m^2}\right)
                      \left(g^{\nu\sigma}-\frac{p^{\nu}p^{\sigma}}{m^2}\right)
        + \frac{1}{2} \left(g^{\mu\sigma}-\frac{p^{\mu}p^{\sigma}}{m^2}\right)
                      \left(g^{\nu\rho}-\frac{p^{\nu}p^{\rho}}{m^2}\right) \\
        - \frac{1}{3} \left(g^{\mu\nu}-\frac{p^{\mu}p^{\nu}}{m^2}\right)
                      \left(g^{\rho\sigma}-\frac{p^{\rho}p^{\sigma}}{m^2}\right)
    \end{multline}
    \end{subequations}
\end{itemize}


\subsection{Vertices}

\begin{table}
  \begin{center}
    {\scriptsize
    \begin{tabular}{l|l}
       $\bar\psi(g_V\gamma^\mu - g_A\gamma^\mu\gamma_5)\psi$
         & $\text{\texttt{va\_ff}}(g_V,g_A,\bar\psi,\psi)$ \\
       $g_V\bar\psi\gamma^\mu\psi$
         & $\text{\texttt{v\_ff}}(g_V,\bar\psi,\psi)$ \\
       $g_A\bar\psi\gamma_5\gamma^\mu\psi$
         & $\text{\texttt{a\_ff}}(g_A,\bar\psi,\psi)$ \\
       $g_L\bar\psi\gamma^\mu(1-\gamma_5)\psi$
         & $\text{\texttt{vl\_ff}}(g_L,\bar\psi,\psi)$ \\
       $g_R\bar\psi\gamma^\mu(1+\gamma_5)\psi$
         & $\text{\texttt{vr\_ff}}(g_R,\bar\psi,\psi)$ \\\hline
       $\fmslash{V}(g_V - g_A\gamma_5)\psi$
         & $\text{\texttt{f\_vaf}}(g_V,g_A,V,\psi)$ \\
       $g_V\fmslash{V}\psi$
         & $\text{\texttt{f\_vf}}(g_V,V,\psi)$ \\
       $g_A\gamma_5\fmslash{V}\psi$
         & $\text{\texttt{f\_af}}(g_A,V,\psi)$ \\
       $g_L\fmslash{V}(1-\gamma_5)\psi$
         & $\text{\texttt{f\_vlf}}(g_L,V,\psi)$ \\
       $g_R\fmslash{V}(1+\gamma_5)\psi$
         & $\text{\texttt{f\_vrf}}(g_R,V,\psi)$ \\\hline
       $\bar\psi\fmslash{V}(g_V - g_A\gamma_5)$
         & $\text{\texttt{f\_fva}}(g_V,g_A,\bar\psi,V)$ \\
       $g_V\bar\psi\fmslash{V}$
         & $\text{\texttt{f\_fv}}(g_V,\bar\psi,V)$ \\
       $g_A\bar\psi\gamma_5\fmslash{V}$
         & $\text{\texttt{f\_fa}}(g_A,\bar\psi,V)$ \\
       $g_L\bar\psi\fmslash{V}(1-\gamma_5)$
         & $\text{\texttt{f\_fvl}}(g_L,\bar\psi,V)$ \\
       $g_R\bar\psi\fmslash{V}(1+\gamma_5)$
         & $\text{\texttt{f\_fvr}}(g_R,\bar\psi,V)$
    \end{tabular} \qquad 
    \begin{tabular}{l|l}
       $\bar\psi(g_S + g_P\gamma_5)\psi$
         & $\text{\texttt{sp\_ff}}(g_S,g_P,\bar\psi,\psi)$ \\
       $g_S\bar\psi\psi$
         & $\text{\texttt{s\_ff}}(g_S,\bar\psi,\psi)$ \\
       $g_P\bar\psi\gamma_5\psi$
         & $\text{\texttt{p\_ff}}(g_P,\bar\psi,\psi)$ \\
       $g_L\bar\psi(1-\gamma_5)\psi$
         & $\text{\texttt{sl\_ff}}(g_L,\bar\psi,\psi)$ \\
       $g_R\bar\psi(1+\gamma_5)\psi$
         & $\text{\texttt{sr\_ff}}(g_R,\bar\psi,\psi)$ \\\hline
       $\phi(g_S + g_P\gamma_5)\psi$
         & $\text{\texttt{f\_spf}}(g_S,g_P,\phi,\psi)$ \\
       $g_S\phi\psi$
         & $\text{\texttt{f\_sf}}(g_S,\phi,\psi)$ \\
       $g_P\phi\gamma_5\psi$
         & $\text{\texttt{f\_pf}}(g_P,\phi,\psi)$ \\
       $g_L\phi(1-\gamma_5)\psi$
         & $\text{\texttt{f\_slf}}(g_L,\phi,\psi)$ \\
       $g_R\phi(1+\gamma_5)\psi$
         & $\text{\texttt{f\_srf}}(g_R,\phi,\psi)$ \\\hline
       $\bar\psi\phi(g_S + g_P\gamma_5)$
         & $\text{\texttt{f\_fsp}}(g_S,g_P,\bar\psi,\phi)$ \\
       $g_S\bar\psi\phi$
         & $\text{\texttt{f\_fs}}(g_S,\bar\psi,\phi)$ \\
       $g_P\bar\psi\phi\gamma_5$
         & $\text{\texttt{f\_fp}}(g_P,\bar\psi,\phi)$ \\
       $g_L\bar\psi\phi(1-\gamma_5)$
         & $\text{\texttt{f\_fsl}}(g_L,\bar\psi,\phi)$ \\
       $g_R\bar\psi\phi(1+\gamma_5)$
         & $\text{\texttt{f\_fsr}}(g_R,\bar\psi,\phi)$
    \end{tabular}}
  \end{center}
  \caption{\label{tab:fermionic-currents}
    Mnemonically abbreviated names of \FORTRAN/ functions implementing
    fermionic vector and axial currents on the left, scalar and
    pseudoscalar currents on the right.}
\end{table}

For fermionic vertices we use the following chiral representation used
in \HELAS/~\cite{HELAS}:
\begin{subequations}
\begin{align}
  & \gamma^0 = \begin{pmatrix} 0 & \mathbf{1} \\ \mathbf{1} & 0
    \end{pmatrix},\;
  \gamma^i = \begin{pmatrix} 0 & \sigma^i \\ -\sigma^i & 0 \end{pmatrix},\;
  \gamma_5 = i\gamma^0\gamma^1\gamma^2\gamma^3
           = \begin{pmatrix} -\mathbf{1} & 0 \\ 0 & \mathbf{1}
           \end{pmatrix}, \\ &
  C = \begin{pmatrix} \epsilon & 0 \\ 0 & - \epsilon \end{pmatrix}
  \; , \qquad \epsilon = \begin{pmatrix} 0 & 1 \\ -1 & 0 \end{pmatrix}  .
\end{align}
\end{subequations}

The conventions for the fermionic vertices are shown in
Table~\ref{tab:fermionic-currents}.

\section{Sample Matrix Element Code}
\label{sec:sample-code}

Here, we present an example for the process-specific \FORTRAN/ code
generated by \OMEGA/, which is compiled and linked by the main
\WHIZARD/ executable.  The process is $e^+e^-\to\mu^+\mu^-$ in the SM.
In this example, the code is equivalent to the sum of Feynman
diagrams, as there are no common subexpressions.  The decomposition
into numbered subroutines is effective in speeding up compilation,
which becomes relevant for large process codes.
\begin{verbatim}
  subroutine calculate_amplitudes (amp, k, mask)
    complex(kind=default), dimension(:,:,:), intent(out) :: amp
    real(kind=default), dimension(0:3,*), intent(in) :: k
    logical, dimension(:), intent(in) :: mask
    integer, dimension(n_prt) :: s
    integer :: h
    p1 = - k(:,1) ! incoming
    p2 = - k(:,2) ! incoming
    p3 =   k(:,3) ! outgoing
    p4 =   k(:,4) ! outgoing
    p12 = p1 + p2
    amp = 0
    do h = 1, n_hel
     if (mask(h)) then
      s = table_spin_states(:,h)
      owf_l1b_1 = vbar (mass(11), - p1, s(1))
      owf_l1_2 = u (mass(11), - p2, s(2))
      owf_l2_3 = v (mass(13), p3, s(3))
      owf_l2b_4 = ubar (mass(13), p4, s(4))
      call compute_fusions_0001 ()
      call compute_brakets_0001 ()
      amp(1,h,1) = oks_l1bl1l2bl2
     end if
    end do
  end subroutine calculate_amplitudes
  subroutine compute_fusions_0001 ()
      owf_a_12 = pr_feynman(p12, + v_ff(qlep,owf_l1b_1,owf_l1_2))
      owf_z_12 = pr_unitarity(p12,mass(23),wd_tl(p12,width(23)), &
         + va_ff(gnclep(1),gnclep(2),owf_l1b_1,owf_l1_2))
  end subroutine compute_fusions_0001
  subroutine compute_brakets_0001 ()
      oks_l1bl1l2bl2 = 0
      oks_l1bl1l2bl2 = oks_l1bl1l2bl2 + owf_z_12*( &
         + va_ff(gnclep(1),gnclep(2),owf_l2b_4,owf_l2_3))
      oks_l1bl1l2bl2 = oks_l1bl1l2bl2 + owf_a_12*( &
         + v_ff(qlep,owf_l2b_4,owf_l2_3))
      oks_l1bl1l2bl2 = - oks_l1bl1l2bl2 ! 2 vertices, 1 propagators
      ! unit symmetry factor
  end subroutine compute_brakets_0001
\end{verbatim}


\end{document}